\documentclass[aps,prx,twocolumn,showpacs,psfig,superscriptaddress,longbibliography]{revtex4-2}
\usepackage{times}
\usepackage{graphicx}
\usepackage{float}
\usepackage{amsmath}
\usepackage{latexsym,amsmath,amssymb,bm,euscript}
\usepackage{color}
\usepackage{subfigure}
\usepackage{epstopdf}
\usepackage[colorlinks=true,linkcolor=blue,citecolor=blue]{hyperref}
\usepackage{soul}
\usepackage[normalem]{ulem}
\usepackage{mathrsfs}
\usepackage{amsmath}
\usepackage{lettrine}
\usepackage{bbding}
\usepackage{xspace}
\usepackage{textcomp}
\usepackage{textcase}
\usepackage{setspace}
\usepackage{mathtools}
\usepackage[ruled, algo2e, linesnumbered]{algorithm2e}
\usepackage{url}
\usepackage{lineno}

\def\para{\ensuremath{/\kern -0.8em /}\xspace}
\def\beqn{\begin{eqnarray}}
\def\eeqn{\end{eqnarray}}
\def\beq{\begin{equation}}
\def\eeq{\end{equation}}

\newcommand{\Beq}{\begin{eqnarray*} }
\newcommand{\Eeq}{\end{eqnarray*} }
\newcommand{\Bmat}{\left(\begin{matrix}}
\newcommand{\Emat}{\end{matrix}\right)}

\graphicspath{{./Fig/}}

\begin{document}
\title{Spin supersolidity in nearly ideal easy-axis triangular quantum antiferromagnet Na$_2$BaCo(PO$_4$)$_2$}

\author{Yuan Gao}
\affiliation{School of Physics, Beihang University, Beijing 100191, China}
\affiliation{CAS Key Laboratory of Theoretical Physics, Institute of 
Theoretical Physics, Chinese Academy of Sciences, Beijing 100190, China}

\author{Yu-Chen Fan}
\affiliation{Institute of Physics, Chinese Academy of Sciences, Beijing 
100190, China}

\author{Han Li}
\affiliation{School of Physics, Beihang University, Beijing 100191, China}
\affiliation{CAS Key Laboratory of Theoretical Physics, Institute of 
Theoretical Physics, Chinese Academy of Sciences, Beijing 100190, China}

\author{Fan Yang}
\affiliation{School of Physics, Beihang University, Beijing 100191, China}

\author{Xu-Tao Zeng}
\affiliation{School of Physics, Beihang University, Beijing 100191, China}

\author{Xian-Lei Sheng}
\affiliation{School of Physics, Beihang University, Beijing 100191, China}
\affiliation{Peng Huanwu Collaborative Center for Research and Education, Beihang University, Beijing 100191, China}

\author{Ruidan Zhong}
\affiliation{Tsung-Dao Lee Institute, Shanghai Jiao Tong University, 
Shanghai 200240, China}
\affiliation{School of Physics and Astronomy, Shanghai Jiao Tong University, Shanghai 200240, China}

\author{Yang Qi}
\affiliation{State Key Laboratory of Surface Physics, Fudan University, 
Shanghai 200433, China}
\affiliation{Center for Field Theory and Particle Physics, Department 
of Physics, Fudan University, Shanghai 200433, China}

\author{Yuan Wan}
\email{yuan.wan@iphy.ac.cn}
\affiliation{Institute of Physics, Chinese Academy of Sciences, Beijing 100190, China}
\affiliation{Songshan Lake Materials Laboratory, Dongguan, Guangdong 523808, China}

\author{Wei Li}
\email{w.li@itp.ac.cn}
\affiliation{CAS Key Laboratory of Theoretical Physics, Institute of 
Theoretical Physics, Chinese Academy of Sciences, Beijing 100190, China}
\affiliation{Peng Huanwu Collaborative Center for Research and Education, Beihang University, Beijing 100191, China}

%=====================================
\begin{abstract} 
Prototypical models and their material incarnations are cornerstones 
to the understanding of quantum magnetism. Here we show theoretically 
that the recently synthesized magnetic compound Na$_2$BaCo(PO$_4$)$_2$ 
(NBCP) is a rare, nearly ideal material realization of the $S=1/2$ triangular-lattice 
antiferromagnet with significant easy-axis spin exchange anisotropy. 
By combining the automatic parameter searching and tensor-network 
simulations, we establish a microscopic model description of this material 
with realistic model parameters, which can not only fit well the experimental 
thermodynamic data but also reproduce the measured magnetization curves 
without further adjustment of parameters. {According to the established model,} the NBCP hosts a spin supersolid state that breaks both the lattice 
translation symmetry and the spin rotational symmetry. Such a state is a spin 
analogue of the long-sought supersolid state, thought to exist in solid Helium 
and optical lattice systems, and share similar traits. The NBCP therefore 
represents an ideal material-based platform to explore the physics of supersolidity 
as well as its quantum and thermal melting.
\end{abstract}
\date{\today}
\maketitle

\noindent{\bf{Introduction}} \\ 
{Quantum magnets are fertile ground for unconventional quantum phases 
and phase transitions.} A prominent example is the $S=1/2$ triangular lattice 
antiferromagnet (TLAF). Crucial to the conception of the quantum spin liquid 
state~\cite{Anderson1973}, its inherent geometric frustration and strong 
quantum fluctuations give rise to exceedingly rich physics. In the presence 
of an external magnetic field, spin anisotropy, and/or spatial anisotropy, 
the system exhibits a cornucopia of magnetic orders and phase transitions
\cite{Chubukov1991,Collins1997,Starykh2015}. In particular, introducing 
an easy-axis spin exchange anisotropy to the TLAF results in the 
\emph{spin supersolid}~\cite{Wessel2005,Melko2005,Heidarian2005,
Prokofev2005,Heidarian2010,WangF2009,Jiang2009} in zero magnetic 
field. Applying a magnetic field along the easy-axis drives the system 
through a sequence of quantum phase transitions by which the spin 
supersolidity disappears and then reemerges~\cite{Yamamoto2014}, 
whereas applying the field in the perpendicular direction yields distinct, 
even richer behaviors~\cite{Yamamoto2019PhaseDiag}. 
The $S=1/2$ easy-axis TLAF therefore constitutes a special platform 
for exploring intriguing quantum phases and quantum phase transitions.

%% ====== Fig 1 ====== %
\begin{figure*}[t]
\includegraphics[width=\textwidth]{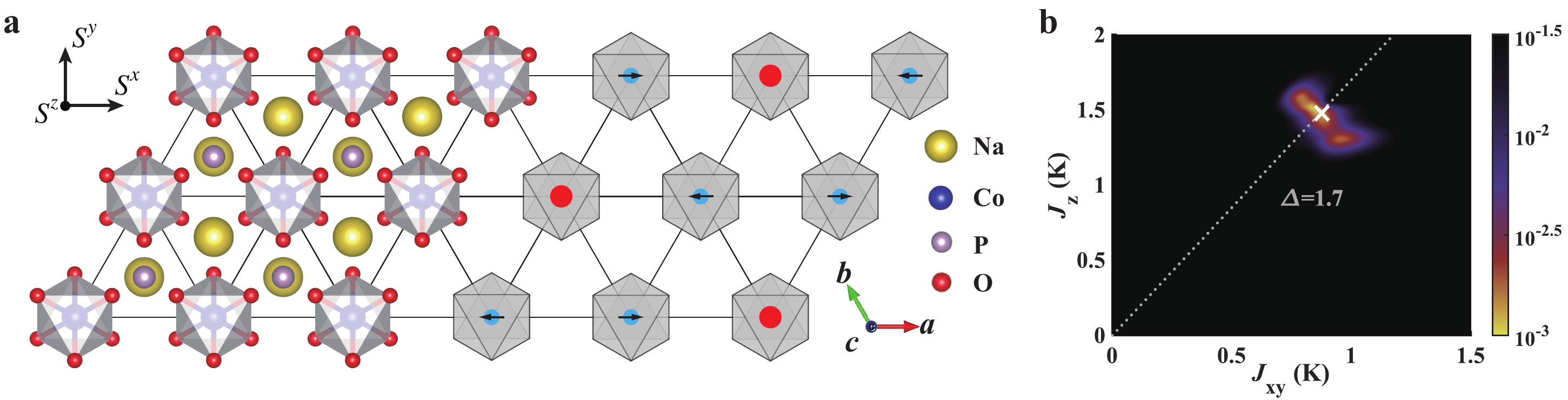}
\renewcommand{\figurename}{\textbf{Fig.}}
\caption{\textbf{Crystallographic structure of NBCP and fitting loss landscape.}
\textbf{a} Top view of the crystallographic structure of the NBCP. Each Co$^{2+}$ 
ion carries an effective $S=1/2$ spin owing to the octahedral crystal environment 
and the spin-orbital coupling. The spins form a triangular network. Its magnetic 
ground state is a spin supersolid: the red (blue) circles represent the positive 
(negative) $S^z$ component, and the black arrows show the direction of the 
$S^{x,y}$ component. The crystallographic ${a}$-, ${b}$-, and ${c}$-axes 
and the spin frame $S^{x,y,z}$ are shown in the bottom right and the top left 
insets, respectively. 
\textbf{b} shows the projection of the fitting loss function onto the 
$J_{xy}$-$J_{z}$ plane (with $J_{\rm PD, \Gamma}$ fixed as zero), 
where brighter color represents less fitting loss and therefore better fit. 
The white cross marks the optimal parameter set, which corresponds 
to an $S=1/2$ TLAF with anisotropy parameter 
$\Delta = J_{z}/J_{xy} \simeq 1.7$ (dashed line). 
}
\label{Fig1}
\end{figure*}
%% ================= %

Lately, a cobalt-based compound Na$_2$BaCo(PO$_4$)$_2$ (NBCP) 
has been brought to light~\cite{Zhong2019,Li2020,Lee2021,Wellm2021}. 
This material features an ideal triangular lattice of Co$^{2+}$ ions, 
each carrying an effective $S=1/2$ spin owing to the crystal field 
environment and the significant spin-orbital coupling~\cite{Liu2018,Wellm2021} 
(Fig.~\ref{Fig1}\textbf{a}). Early thermodynamic measurements show that 
NBCP does not order down to $\sim 300$~mK with a large magnetic 
entropy ($\sim 2$~J~mol$^{-1}$~K$^{-1}$) hidden below 
that temperature scale~\cite{Zhong2019}. A later thermodynamic 
measurement reveals a specific heat peak at $\sim150$~mK, 
which accounts for the missing entropy and points to a possible 
magnetic ordering in zero magnetic field~\cite{Li2020}. However, 
the muon spin resonance ($\mu$SR) experiment finds strong 
dynamical fluctuation down to 80~mK~\cite{Lee2021} which may 
suggest a spin-liquid like state. The multitude of experimental results 
call for a theoretical assessment.

Previous works have attempted at establishing the spin exchange interactions in this compound. 
The authors in Ref.~\onlinecite{Li2020} suggest an exchange coupling $\sim 2$~K based on an analysis of the magnetic susceptibility data, which is an order of magnitude smaller than an earlier estimate of 21.4~K in Ref.~\onlinecite{Zhong2019}. Meanwhile, a first-principle calculation suggests potentially significant Kitaev-type exchange interaction~\cite{Wellm2021}. Despite these efforts, the precise spin Hamiltonian, its magnetic ground states, as well as the connection to experimental data, are yet to be established.

In this work, we show theoretically that NBCP can be well-described by 
a $S=1/2$ easy-axis TLAF with negligible perturbations. We establish 
the microscopic description of NBCP with realistic model parameters 
by fitting the model to intermediate- and high-temperature experimental 
thermal data. We expedite the fitting process by using the Bayesian 
optimization~\cite{Yu2021} equipped with an efficient quantum many-body 
thermodynamic solver --- exponential tensor renormalization group 
(XTRG)~\cite{Chen2019,Lih2019}. Our model is corroborated by 
reproducing quantitatively the experimental low-temperature 
magnetization curves by density matrix renormalization group 
(DMRG)~\cite{White1992} calculations. Furthermore, 
we are able to put the various experimental results into a coherent 
picture and connect them to the physics of the spin supersolid state. 
Therefore, the NBCP represents a rare material realization of this 
prototypical model system and thereby the spin supersolidity. The 
small exchange energy scale in this material ($\sim 1$ K) implies 
that the phases of the NBCP can be readily tuned by weak or 
moderate magnetic fields. Our results also highlight the strength 
of the many-body computation-based, experimental data-driven 
approach as a methodology for studying quantum magnets.

\noindent{\bf{Results}}\\
% ===== Compound NBCP & Anisotropic Hamiltonian ===== %
\textbf{Crystal symmetry and the spin-1/2 model.} 
Figure~\ref{Fig1}\textbf{a} shows the lattice structure of NBCP and the 
crystallographic $a$-, $b$-, and $c$-axes. Due to the octahedral crystal 
field environment 
and the spin-orbital coupling, each Co$^{2+}$ ion forms an effective 
$S=1/2$ doublet in the ground state, which is separated from higher 
energy multiplets by a gap of $\sim71$~meV (see Supplementary 
Note~1). Super-super-exchange path through two intermediate oxygen 
ions produces exchange interactions between two nearest-neighbor
(NN) spins, thereby connecting them into a triangular network (see 
density functional theory calculations in the Supplementary Note~2). 
Further neighbor spin exchange interactions are suppressed by the 
long distance. Meanwhile, the inter-layer exchange interactions are 
expected to be much smaller than the intra-layer couplings owing to the 
non-magnetic BaO layer separating the adjacent cobalt layers. Therefore, 
we model the NBCP in the experimentally relevant temperature window as 
a $S=1/2$ TLAF with dominant NN exchange interactions. 
This hypothesis will be justified \textit{a posteriori}. 

The crystal symmetry constrains the NN exchange interactions as follows~\cite{Tinkham1964}. 
The three-fold symmetry axis 
$\parallel  c$ passing through each lattice site relates the 
exchange interactions on the 6 bonds emanating from that site. On a given bond, 
there is a 2-fold symmetry axis passing through that bond and a center of inversion. 
The former symmetry forbids certain components of the off-diagonal symmetric 
exchange interaction, whereas the latter forbids Dzyaloshinskii-Moriya 
interactions. We obtain 
\begin{equation}
H_{ij}=
\sum_{\alpha,\beta} J^{\alpha \beta} S_i^{\alpha} S_j^{\beta},
\label{Eq:TLH}
\end{equation}
with
\begin{equation}
J^{\alpha\beta} = 
\begin{pmatrix}
J_{xy}+2 J_{\rm PD} \cos\varphi
&-2J_{\rm PD} \sin\varphi
&-J_{\Gamma} \sin\varphi \\
-2J_{\rm PD} \sin\varphi
&J_{xy}-2J_{\rm PD} \cos\varphi
& J_{\Gamma} \cos\varphi\\
-J_{\Gamma} \sin\varphi
& J_{\Gamma} \cos\varphi
& J_z
\end{pmatrix}, \nonumber
\end{equation}
where $i,j$ are a pair of neighboring lattice sites, and $\alpha, 
\beta$ label the spin $x,y,z$ components~\cite{Li2015YMGO2,LiYD2016,Zhu2018}.
We choose the spin 
frame such that $x\parallel  a$, $z\parallel  c$. 
$\varphi=\{0,\frac{2\pi}{3},-\frac{2\pi}{3}\}$ for three different 
types of NN bonds parallel to $ a$, $ b$, and 
$-({a}+{b})$, respectively. $J_{xy}$, $J_z$, 
$J_{\Gamma}$, and $J_{\rm PD}$ are respectively the XY, Ising, 
off-diagonal symmetric, and pseudo-dipolar exchange couplings 
({see more details in Supplementary Note 3}). The entire model 
parameter space is thus spanned by the four exchange constants, 
two Land\'e factors ($g_{ab}$ and $g_c$, for perpendicular and 
parallel to the $c$-axis, respectively), as well as two van Vleck 
paramagnetic susceptibilities ($\chi_{ab}^{\rm vv}, \chi_c^{\rm vv}$), 
all of which are taken to be constants in the experimentally relevant 
temperature/magnetic field window.
\\

% ====== Fig 2 ====== %
\begin{figure*}[t]
\includegraphics[width= \textwidth]{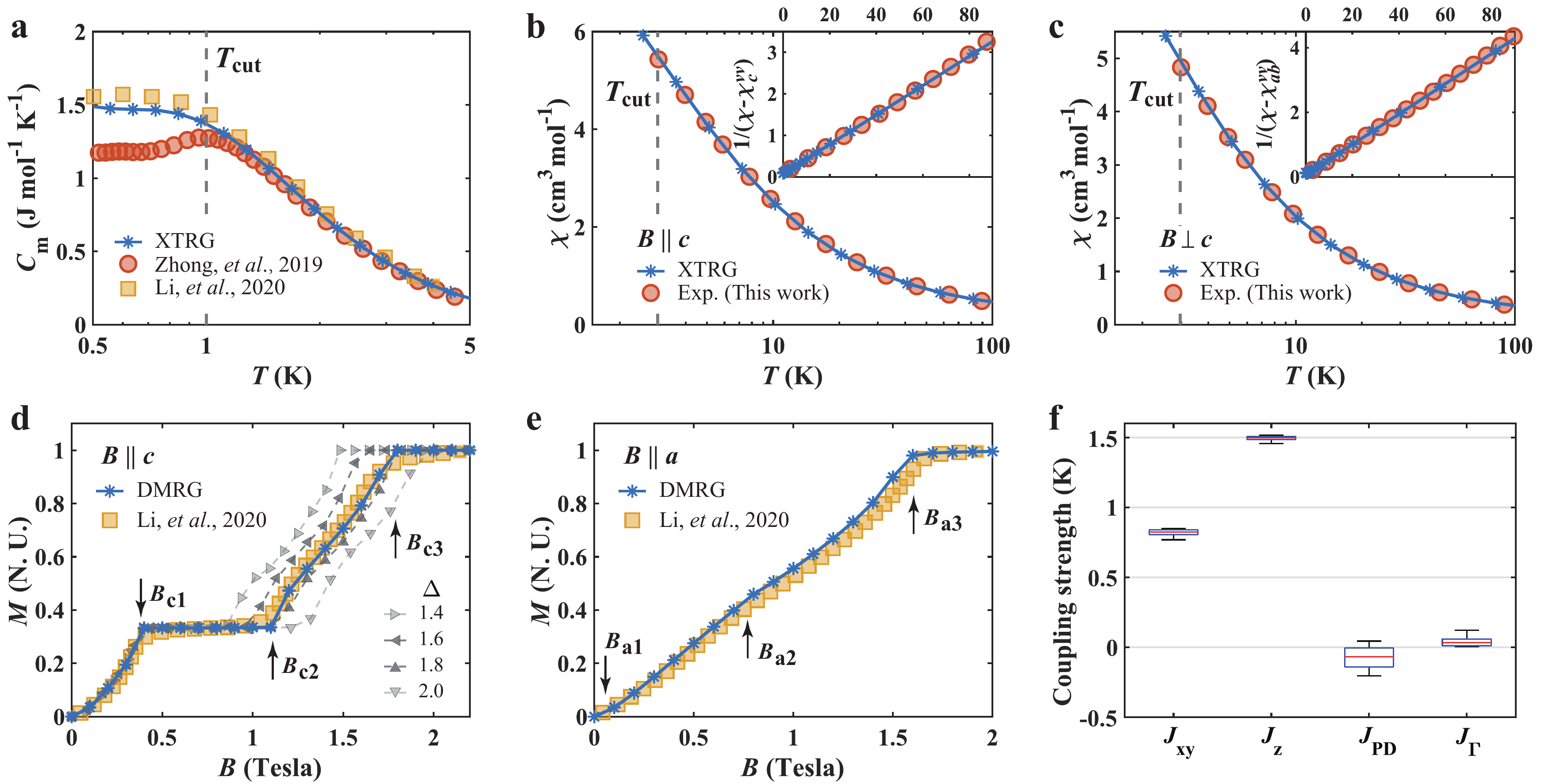}
\renewcommand{\figurename}{\textbf{Fig.}} 
\caption{\textbf{Thermodynamic data fittings, 
magnetization curves, and the optimal model 
parameters.} \textbf{a} shows the magnetic specific 
heat $C_{\rm m}$ of the NBCP as a function of temperature $T$, reported by two 
independent measurements ({\it Zhong et al. 2019}~\cite{Zhong2019} and {\it Li et al. 2020}~\cite{Li2020}) 
as well as the best fit. The gray dashed line marks $T_{\rm cut}$. The 
XTRG fitting is performed for experimental data at $T>T_\mathrm{cut}$. 
\textbf{b,~c} show respectively parallel and perpendicular to the $c$-axis 
magnetic susceptibilities with the experimental data measured in this work 
(see Methods) and the best fit. 
\textbf{d,~e} show the magnetization as a function of the magnetic field in the 
${c}$ and ${a}$ directions, respectively. The experimental data were obtained 
from~{\it Li et al. 2020}~\cite{Li2020} with the van Vleck paramagnetic contributions 
subtracted off. The model prediction is obtained from DMRG, which 
agrees quantitatively with experiment. The grey 
lines show magnetization curves obtained from models with various 
anisotropy parameters $\Delta$, with $J_{xy}$ tuned to the optimal value. 
\textbf{f} shows the standard box plot of the optimal parameters, 
with the red line indicating the median value, the top(bottom) 
edge of the box representing the upper(lower) quartile of the 
best 20 model parameter sets found in 450 Bayesian optimization runs.
}
\label{Fig2}
\end{figure*}
% ================= %

\noindent
\textbf{Determination of the model parameters.} 
We determine the model parameters in Eq.~(\ref{Eq:TLH}) by fitting the 
experimental magnetic specific heat ($C_{\rm m}$) and magnetic susceptibility
($\chi$) data at temperature 
$T \geq T_{\rm cut}$, where $T_{\rm cut}=1$~K for $C_{\rm m}$ and 3~K for $\chi$. 
Note the magnetic susceptibilities are remeasured 
in this work with high quality samples.
For each trial parameter set, we compute the same thermodynamic quantities 
from the model by using the XTRG solver~\cite{Chen2018,Lih2019}. We search 
for the parameter set that minimizes the total loss function through an unbiased 
and efficient Bayesian optimization process~\cite{Yu2021}. See Methods for 
more details.

We set the cutoff temperature $T_\mathrm{cut}$ based on the following considerations. The experimental data from independent measurements 
agree with each other at $T>T_\mathrm{cut}$, and our XTRG solver does 
not exhibit significant finite-size effects above $T_\mathrm{cut}$. Meanwhile,
the $T_\mathrm{cut}$ has to be less than or comparable with the characteristic 
energy scale of the material. These constraints fix 
$T_\mathrm{cut}$ to our present choices.

The searching process yields the following optimal parameter set: 
$J_{xy}=0.88$~K, $J_{z}=1.48$~K, and $J_{\Gamma, \rm PD}$ are negligible. 
The Land\'{e} factors $g_{ab}=4.24$, and $g_c = 4.89$. The van Vleck susceptibilities $\chi^{\rm vv}_{ab}=0.149$ cm$^3$ mol$^{-1}$
and $\chi^{\rm vv}_c=0.186$ cm$^3$ mol$^{-1}$.
To ensure that the algorithm does converge to the global minimum, we 
project the loss function onto the $(J_{xy},J_z)$ plane in Fig.~\ref{Fig1}\textbf{b}, where, for fixing values of $J_{xy},J_z$, the loss function is minimized over 
the remaining parameters. The fitting landscape reveals a single minimum. 
The estimated value of exchange parameters and their bounds of uncertainty 
are shown in Fig.~\ref{Fig2}\textbf{f}. 

The small uncertainties in $J_{xy}$ and $J_z$, as well as the small 
loss, indicate that the experimental data are well captured by our 
parameters. Indeed, Fig.~\ref{Fig2}$\textbf{a-c}$ show respectively 
the specific heat and the magnetic susceptibility as functions of 
temperature. We find excellent agreement between the model calculations 
and the experiments within the fitting temperature range $T \geq T_{\rm cut}$. Reassuringly, the Land\'e factors obtained by us are in excellent 
agreement with the latest electron-spin resonance measurement 
($g_{ab} = 4.24$ and $g_c = 4.83$)~\cite{Wellm2021}. 
We have also calculated the magnetic specific heat $C_{\rm m}$ in 
non-zero magnetic fields and find good agreement with the 
experiments whenever the two independent measurements 
\cite{Zhong2019,Li2020} mutually agree (Supplementary Note 4). 
Note in Fig.~\ref{Fig2}\textbf{a} the two experimental data sets of 
specific heat differ at $T<T_{\rm cut}$. 
Our model's behavior below $T_{\rm cut}$ is in agreement with one 
of them. The discrepancy in the experimental data calls for further 
investigation.

Our model parameters pinpoint to an almost ideal $S=1/2$ TLAF with 
significant easy-axis anisotropy $\Delta = J_z/J_{xy}\approx 1.68$. 
In particular, the negligible off-diagonal exchange interactions imply 
that the NBCP features an approximate U(1) spin rotational symmetry 
with respect to $c$ axis. As a result, the magnetization curve with field
$B \parallel  c$ in Fig.~\ref{Fig2}\textbf{d} has a couple of idiosyncratic 
features: It shows a $1/3$-magnetization plateau in an intermediate 
field range $[B_{c1},B_{c2}]$, and another fully magnetized plateau 
above the saturation field $B_{c3}$. 
As an independent estimate of $\Delta$, we note the semi-classical 
analysis shows that $B_{c2}/B_{c1} = \Delta - 1/2+\sqrt{\Delta^2+\Delta-7/4}$ 
and $B_{c3}/B_{c1} = 2\Delta + 1$ (Supplementary Note 5). 
Using the experimental values $B_{c1} \approx 0.35$~T and $B_{c3}\approx1.62$~T~\cite{Li2020}, we estimate $\Delta \approx 1.81$, 
which is consistent with the Bayesian search result. Meanwhile, 
using these numbers, we can estimate $B_{c2} \approx 1.10$~T, 
which is fairly close to the experimental value of $1.16$~T~\cite{Li2020}. 

As a corroboration of our model, we perform DMRG calculations 
of the zero-temperature magnetization curves and find quantitative 
agreement with the experimental results (Fig.~\ref{Fig2}\textbf{d,~e}). 
With no further adjustment of the parameters, the model can not only 
produce the correct transition fields but also the details of the magnetization 
curve between the transitions. We find that the magnetization curve 
$\parallel  c$ is a sensitive diagnostic for the anisotropy parameter 
$\Delta$. The agreement between the model and the experiments is 
quickly lost when $\Delta$ deviates slightly from the optimal value  
in Fig.~\ref{Fig2}\textbf{d}. 

Taken together, the broad agreements between the semi-classical 
estimates, the quantum many-body calculations, and experimental 
data strongly support the $S=1/2$ easy-axis TLAF as an effective 
model description for the NBCP.
\\

% ====== Fig 3 ====== %
\begin{figure}[t]
\includegraphics[width = 0.95 \linewidth]{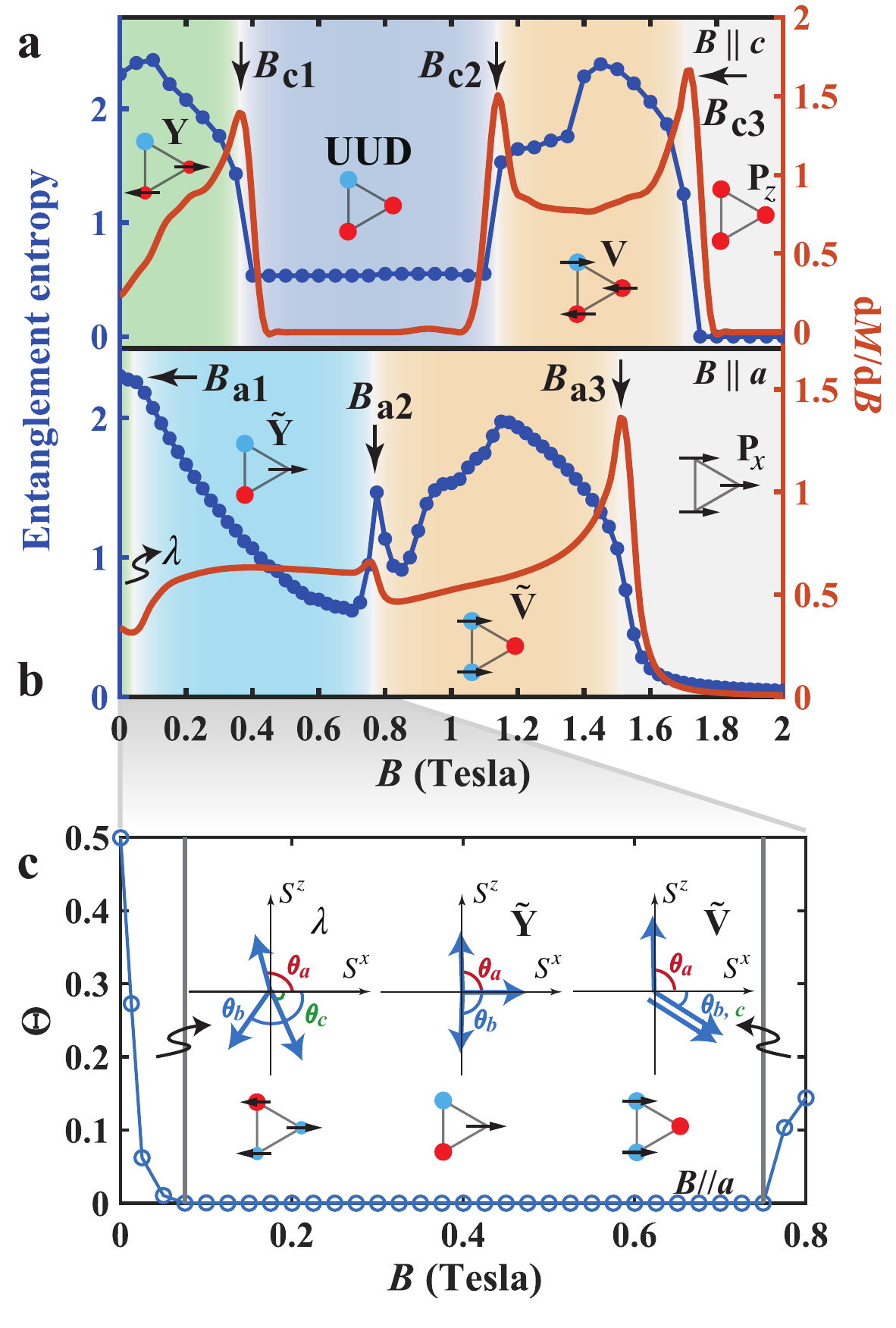}
\renewcommand{\figurename}{\textbf{Fig.}}
\caption{\textbf{Zero-temperature phase diagrams under in- and out-of-plane 
fields.} \textbf{a} shows the entanglement entropy and differential 
magnetic susceptibility $dM/dB$ as functions of field $B$ in the ${c}$ direction. 
The Y, UUD, V, and P$_z$ phases are shaded in different colors. $B_{c1,c2,c3}$ are the three critical fields separating these phases. The corresponding spin configurations are also shown. Red (blue) circles represent the positive (negative) $S^z$ component. Black arrows show the direction of the in-plane ($S^{x,y}$) component. \textbf{b} is similar to \textbf{a} but for  field $B\parallel  a$. 
The $\lambda$ phase, $\widetilde{\mathrm{Y}}$ phase, $\tilde{\rm V}$ phase, 
and P$_x$ phase are separated by three critical fields $B_{a1,a2,a3}$. \textbf{c} is a zoom-in of \textbf{b}, showing the side view of the spin configurations, as well as the evolution of the order parameter $\Theta$ as a function of the field (see main text for definition).}
\label{Fig3}
\end{figure}
% ===================== %

\noindent
\textbf{Field-tuning of the spin supersolid state.} 
The $S=1/2$ easy-axis TLAF exhibits a sequence of magnetic 
phases and quantum phase transitions driven by magnetic fields~\cite{Yamamoto2014,Yamamoto2015,Sellmann2015}. 
The experimentally measured differential susceptibilities ($dM/dB$) 
show a few anomalies when the field is $\parallel  c$ and 
$\parallel  a$ and are attributed to quantum phase transitions
\cite{Li2020}. Here, we clarify the nature of the magnetic orders 
of NBCP based on the $S=1/2$ easy-axis TLAF model.

Figure~\ref{Fig3}\textbf{a} shows the theoretical zero-temperature phase 
diagram of our model in field $\parallel c$, obtained from DMRG calculations. 
The system goes through successively the Y, Up-Up-Down (UUD), V, 
and the polarized (P$_z$) phases with increasing field. These phases 
are separated by three critical fields $B_{c1,c2,c3}${ = 0.36 {\rm T}, 
1.14 {\rm T}, 1.71 {\rm T}} discussed in the previous section, 
which are manifested as peaks in $dM/dB$. 
The Y phase, as well as the V phase, spontaneously break both the lattice 
translation symmetry and the U(1) symmetry, thereby constituting the supersolid 
state analogous to that of the Bose atoms. The UUD phase, on the other hand, 
restores the U(1) symmetry but breaks the lattice translation symmetry. This 
state is analogous to a Bose Mott insulator state. The magnetic plateau 
associated with the UUD state reflects the incompressibility of the Mott insulator. 

% ====== Fig 4 ====== %
\begin{figure*}[t]
\includegraphics[width = 0.8 \linewidth]{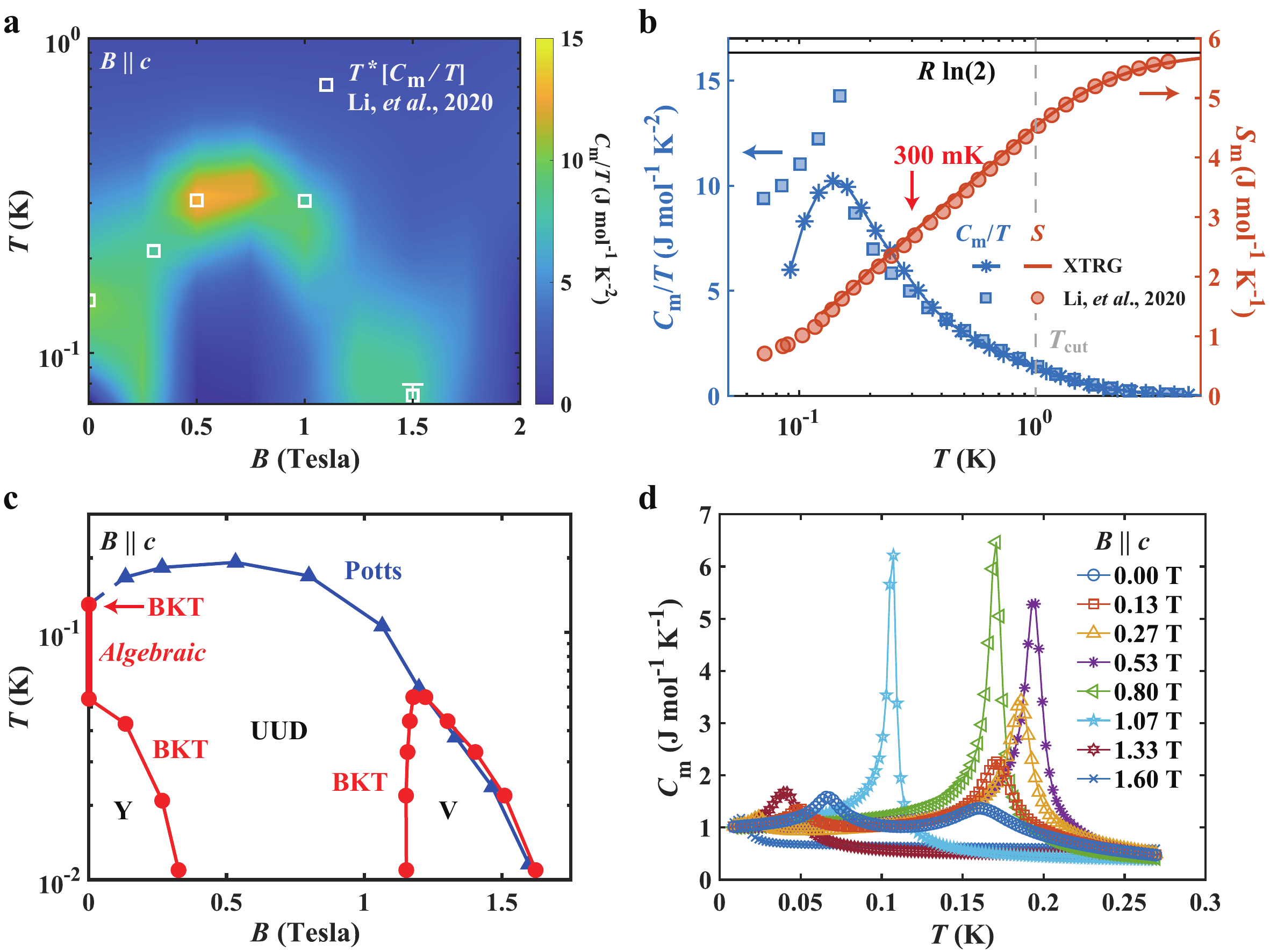}
\renewcommand{\figurename}{\textbf{Fig.}} \caption{
\textbf{Finite-temperature phase diagram of the NBCP under 
magnetic fields $\parallel  c$.} \textbf{a} shows $C_{\text{m}}/T$, 
calculated from XTRG, as a function of temperature $T$ and 
magnetic field $\parallel  c$. The open squares mark the peaks 
of the experimental $C_{\text{m}}/T$ data ({\it Li et al. 2020}~\cite{Li2020}). 
\textbf{b} shows $C_{\rm m}/T$ under zero field, and the magnetic entropy 
$S_{\rm m}$, as functions of temperature. The experimental $S_{\rm m}$ 
data are obtained by integrating numerically the $C_{\rm m}/T$ data. 
The experimental entropy curve is shifted by $0.55~ 
\text{J}~\text{mol}^{-1}~\text{K}^{-1}$ to match the saturation 
value $R \ln(2)$ at sufficiently high temperature.
\textbf{c} shows the  temperature-field phase diagram constructed from 
the classical Monte Carlo (MC) simulation. The blue triangles mark the 
three-state Potts transitions determined by using the Binder cumulants. 
Red solid circles mark the BKT transitions determined from the spin 
stiffness (see Methods). The zero field BKT transition temperatures are 
reported by Ref.~\cite{Stephan2000}. {The red solid line along the 
vertical axis under zero field indicates the regime with algebraic spin 
correlation in the $S^z$ spin component.} \textbf{d} shows the specific 
heat as a function of temperature for various representative magnetic 
field $\parallel c$, obtained from the classical MC simulation.
}
\label{Fig4}
\end{figure*}

The situation is yet more intricate when the field $\parallel  a$. 
DMRG calculations show that the system goes through the $\lambda$, 
$\tilde{\rm Y}$, $\tilde{ \rm V}$ (see Fig.~\ref{Fig3}\textbf{c} for 
an illustration), and the quasi-polarized ${\rm P}_x$ phases, 
which are separated by three critical fields $B_{a1}${ = 0.075~{\rm T}}, $B_{a2}${ = 0.75~{\rm T}}, 
and $B_{a3}${ = 1.51~{\rm T}}. The presence of a field $\parallel a$ 
breaks the U(1)-rotational symmetry with respect to the $S^z$ axis but 
preserves the $\pi$-rotational ($Z_2$) symmetry with respect to the $S^x$ 
axis. The $\lambda$ phase, where the spins sitting on three magnetic 
sublattices form the greek letter ``$\lambda$", spontaneously breaks 
both the $Z_2$ symmetry and the lattice translation symmetry. The $Z_2$ 
symmetry is restored in the $\tilde{\rm Y}$ phase, and spontaneously 
broken again in the $\tilde{\rm V}$ phase. 

Comparing to the field-induced transitions in $B \parallel  c$ case, 
the transitions at $B_{a1,a2}$ show much weaker anomalies in $dM/dB$ 
when $B \parallel {a}$. Numerically, we detect these two transitions using 
an order parameter $\Theta=\frac{1}{\pi} |(\theta_a - \theta_b - \theta_c)|$, 
where $\theta_{a,b,c}$ measure the angle between the spin moments 
on three sublattice and the $S^x$ axis. $\Theta=0$ when the $Z_2$ 
symmetry is respected and $\Theta \neq0$ when it is broken. 
Figure.~\ref{Fig3}\textbf{c} shows $\Theta$ as a function of field, 
from which we can delineate the boundaries between $\lambda$, 
$\tilde{\rm Y}$, and $\tilde{\rm V}$ phases, with the critical fields 
$B_{a1}\approx 0.07$~T and $B_{a2}\approx 0.75$~T. The small value 
of $B_{a1}$ implies it could be easily missed in experiments. Meanwhile, 
the weak anomalies in $dM/dB$ associated with $B_{a1,a2}$ make 
them difficult to detect in thermodynamic measurements. We note 
that $dM/dB$ shows a broad peak in the $\tilde{\rm Y}$ phase 
at $\sim$~0.4-0.5~T~\cite{Li2020}. This peak appears in 
the experimental data and was previously interpreted as 
a transition. The true transitions ($B_{a1,a2}$) are in fact 
above and/or below the said peak. We also note that, despite 
of the weak anomaly observed numerically at $B_{a2}$, 
the quantum phase transition there is likely of first-order 
from symmetry analysis: the $\tilde {\rm Y}$ and $\tilde {\rm V}$ 
phases both have a 6-fold ground-state degeneracy and 
they have incompatible symmetry breaking; thus the transition 
cannot be continuous according to Landau's paradigm. \\

\noindent
\textbf{Strong spin fluctuations and phase diagram at finite temperature.} 
Having established the zero-temperature phase diagram of the NBCP, 
we now move on to its physics at finite temperature. Figure~\ref{Fig4}\textbf{a} 
shows the contour plot of $C_{\rm m}/T$ as a function of temperature and field 
$B \parallel  c$. In the temperature window accessible to the XTRG, 
we find a broad peak near zero field, which moves to higher temperature 
and becomes sharper as field increases. These features are in qualitative 
agreement with the experimental findings.

Figure~\ref{Fig4}\textbf{b} shows a cross-section of the contour plot at 
zero field. The model produces a peak in $C_{\rm m}/T$ at $\approx 150$
mK, which is in excellent agreement with the experimental data from 
Ref.~\cite{Li2020}. Note this temperature is well below the temperature
window (above $T_{\rm cut}$) used for fitting, {the difference between
theory and experiment at very low temperatures may be ascribed to the 
finite-size effect inherent in the XTRG calculations (see Methods)}. The 
magnetic entropy also shows quantitative agreement with the experimental 
data. In particular, there is still a considerable amount of magnetic entropy 
to be released at 300~mK (and even down to 150~mK). The missing entropy 
at 300~mK reported in Ref.~\cite{Zhong2019} can be ascribed to the small 
spin interaction energy scale and high degrees of frustration in the NBCP.

To understand the finite-temperature phase diagram of the NBCP, we perform 
a Monte Carlo (MC) simulation of the classical TLAF model with appropriate 
rescaling of temperature and magnetic field~\cite{Miyashita1985,Stephan2000,
Sheng1992,Shannon2011}. This approximation is amount to 
neglecting fluctuations in the imaginary time direction in the coherent-state 
path integral of the $S=1/2$ TLAF. As the finite-temperature phase transitions 
are driven by thermal fluctuations, we expect that the salient features produced 
by the classical MC simulations are robust. Meanwhile, the MC simulation 
allows for accessing much larger system sizes and lower temperatures 
comparing to the XTRG for quantum model simulations.

The physics of the classical TLAF model is well documented; here, 
we focus on the features that can be directly compared with available 
experimental data. Figure.~\ref{Fig4}\textbf{c} shows the MC-constructed 
$T$-$B$ phase diagram with $B \parallel {c}$. Figure.~\ref{Fig4}\textbf{d} 
shows the specific heat as a function of temperature for various representative
fields. The phase diagram shows a broad dome of UUD phase, beneath 
which lie the Y phase at low field and the V phase at high field. The UUD 
phase and the paramagnetic phase are separated by a transition of three-state 
Potts universality; the Y and V phases and the UUD phase are seperated by the Berezinskii-Kosterlitz-Thoueless (BKT) transitions. Note the MC simulation may seem to suggest the onset of the V phase precedes that of the UUD phase at high field; this is an finite-size effect. On symmetry ground, we expect that the onset of the UUD phase either precedes that of the V phase through two continuous transitions, or the system enters the V phase directly through a first-order transition.

At zero field, the specific heat shows two broad peaks, which are related to the two 
BKT transitions accompanying the onset of the algebraic long-range order in $S^z$ 
and $S^x$ components, respectively~\cite{Miyashita1985,Stephan2000}. The 
experimentally observed specific heat peak $\sim 150$~mK {may} 
be related to the higher-temperature BKT transition; the lower-temperature BKT 
transition {(around 54~mK as estimated by classical 
MC simulations)} is yet to be detected as they lie below 
the temperature window probed by the previous experiments. 
The strong dynamical spin fluctuations found in $\mu$SR experiment at 80~mK 
is naturally attributed to the algebraic long-range order in the $S^z$ component. 
Note there exists arguments for a third BKT transition~\cite{Sheng1992} 
although it is not observed in classical MC simulations~\cite{Stephan2000}.

When the magnetic field is switched on, the specific heat shows a sharp peak signaling the three-state Potts transition from the high temperature paramagnetic phase to the UUD phase, corresponding to onset of the long-range order in the 
$S^z$ component. This is consistent with the experimentally observed sharp 
specific heat peak at finite fields~\cite{Li2020}. At lower temperature, the specific 
heat shows a much weaker peak related to the BKT transition into either the Y 
phase or the V phase, corresponding to the onset of the algebraic long-range 
order in $S^x$. The lower temperature BKT transitions are yet to be detected 
by experiments. \\

\noindent{\bf{Discussion}}\\
% //Summary of main conclusion
The supersolid, a spatially ordered system that exhibits superfluid behavior, 
is a long-pursued quantum state of matter. The question of whether such a 
fascinating phase of matter exists in nature has spurred intense research 
activity, and the search for supersolidity has become a multidisciplinary 
endeavour~\cite{Kim2004,Kim2012,Li2017Nature,Leonard2017,Tanzi2019,
Norcia2021}. The early claim of observation in He-4~\cite{Kim2004} turned out 
to be an experimental artifact~\cite{Kim2012}. Nevertheless, it has inspired new 
lines of research in ultracold quantum gases~\cite{Li2017Nature,Leonard2017,
Tanzi2019,Norcia2021}. Meanwhile, it has been proposed theoretically that 
the ultracold Bose atoms in a triangular optical lattice can host a supersolid 
state~\cite{Wessel2005,Melko2005,Heidarian2005,Prokofev2005}. 
Yet, the realization of such a proposal has not been reported up to date.

An equivalent, yet microscopically different route to the triangular lattice 
supersolidity is via the easy-axis $S=1/2$ TLAF magnet. The spin up/down 
state of  a magnetic ion can be viewed as the occupied/empty state of the 
lattice site by a Bose atom, and the spin rotational symmetry with respect 
to the easy axis is mapped to the U(1) phase rotation symmetry. 
By virtue of this mapping, the spin ground state in the easy-axis TLAF, 
which spontaneously breaks both lattice translation symmetry and spin 
rotational symmetry, is equivalent to the supersolid state of Bose atoms.

{Despite its simple setting, ideal $S=1/2$ TLAF has rarely been found in 
real materials. Although TLAFs with higher spin ($S>1/2$) are known
\cite{Collins1997}, $S=1/2$ systems with equilateral triangular lattice 
geometry, such as Ba$_3$CoSb$_2$O$_9$~\cite{Doi2004,Shirata2012,
Zhou2012,Susuki2013,Ma2016,Sera2016,Ito2017,Kamiya2018} 
and Ba$_8$CoNb$_6$O$_{24}$~\cite{Rawl2017,Cui2018}, 
were synthesized and characterized not until recently. The former 
shows easy-plane anisotropy~\cite{Yamamoto2015,Kamiya2018}, 
whereas the latter material is thought to be nearly spin isotropic
\cite{Rawl2017,Cui2018,Chen2019TLH}. To the best of our knowledge, 
ideal $S=1/2$ easy-axis TLAFs are yet to be found. In this work, 
we} show that the NBCP is an almost ideal material realization 
of such an $S=1/2$ easy-axis TLAF with the anisotropy parameter 
$\Delta \approx 1.7$. 

{Our model arranges the various pieces of available experimental data into a coherent picture by connecting them to the rich physics of the TLAF model. It permits a quantitative fit of the thermodynamic data, including specific heat $C_{\rm m}$ and magnetic susceptibility $\chi$ down to intermediate and even low temperatures. In particular, we obtain the $C_{\rm m}/T$ peak at around 150~mK observed in experiments~\cite{Li2020}, which we associate to the BKT transition. Furthermore, we are able to accurately reproduce the spin state transition fields observed in previous AC susceptibility measurements along both $a$ and $c$ axes and clarify their nature.}

{The obtained spin exchange interactions are on similar orders of magnitude 
as previous estimation based on the Curie-Weiss fitting of the magnetic susceptibility~\cite{Li2020} and first-principle calculation~\cite{Wellm2021}. However, the first-principle calculation suggests a significant Kitaev-type exchange interaction (in a rotated spin frame)~\cite{Wellm2021}, whereas our model, being directly fitted from the experimental data, possesses a nearly ideal U(1) symmetry and negligible Kitaev-type interaction (see more discussions in Supplementary Note 3). The nearly ideal U(1) symmetry in this material is indicated by the well-quantized magnetization plateau (Fig.~\ref{Fig2}d), which would be absent without the U(1) symmetry.}

In our fitting procedure, we have omitted 
at the outset all further-neighbor exchange interactions on the ground that 
their magnitude must be suppressed by the large distance between 
further-neighbor Co$^{2+}$ ions. This can be verified by including in the 
model a second-neighbor spin-isotropic exchange interaction $J_2$.
To verify it, we have performed addtional 400 Bayesian iterations and find 
$J_2$ with the median value $\sim 0.1$~K amongst the best 20 parameter sets, 
which are negligibly small. We thus conclude the obtained optimal parameters 
in the simulations are robust.

Despite the essential challenge in the first-principle calculations of the 
strongly correlated materials, we may nevertheless employ the density 
functional theory (DFT) + U approach to justify certain aspects of the 
microscopic spin model that are accessible to this approach. 
First of all, we find the charge density distributions of 3$d$ electrons 
of Co$^{2+}$ ions are well separated from one triangular plane to another 
(see Supplementary Note~2), which ensures two-dimensionality of 
the compound. Moreover, the in-plane charge density distribution 
reveals clearly a super-super-exchange path between the two NN 
Co$^{2+}$ ions. We construct the Wannier functions of $d$-orbitals 
of Co$^{2+}$ ions and extract the 
hopping amplitude $t$ between two {NN} Co$^{2+}$ ions. 
From the second-order perturbation theory in $t/U$, the 
{NN} exchange coupling can be estimated to be on the order of $2\sim 3$~K 
for moderate and typical $U_{\rm eff}=4\sim6$~eV in this Co-based 
compound~\cite{Wellm2021}, which is consistent with the energy scales of the spin model.

The accurate model for the NBCP also points 
to future directions for the experimentalists to explore. The model hosts 
a very rich phase diagram in both temperature and magnetic field, 
which are yet to be fully uncovered by experiments. In particular, 
the model shows a second BKT transition at $\sim 50$~mK in zero field; 
in finite field $B \parallel  a$, the model shows two subtle transitions at 
$B_{a1}\approx0.07$~T and $B_{a2}\approx 0.75$~T. These transitions 
may be detected by nuclear magnetic resonance~\cite{Hu2020TMGO}, 
magneto-torque measurements~\cite{Modic2020}, and magnetocaloric 
measurements~\cite{Rost2009,Fortune2009CCB,Bachus2020GP}. 
{Neutron scattering experiments can also be employed to detect 
the simultaneous breaking of discrete lattice symmetry and spin U(1) 
rotational symmetry, as well as the behaviors of spin stiffness, 
so as to observe spin supersolidity in this triangular quantum magnet.}
On the theory front, while the $S=1/2$ easy-axis TLAF and its classical 
counterpart share similar features in their finite-temperature phase 
diagrams, it was realized early on that the quantum model also possess 
peculiar traits that are not fully captured by the classical model~\cite{Sheng1992}. 
Clarifying these subtleties in the context of NBCP would also presents 
an interesting problem for the future. \\

\noindent{\bf{Methods}}\\  
\noindent{\textbf{Exponential tensor renormalization group.}}
The thermodynamic quantities including the magnetic specific 
heat $C_{\rm m}$, and magnetic susceptibility $\chi$ can be 
computed with the exponential 
tensor renormalization group (XTRG) method~\cite{Chen2018,
Lih2019}. In practice, we perform XTRG calculations on the 
Y-type cylinders with width $W=6$ and length up to $L=9$ 
(denoted as YC6$\times$9, see Supplementary Note~4), 
and retain up to $D=400$ states with truncation errors 
$\epsilon \lesssim 1\times 10^{-4}$ (down to 1~K) and 
$\lesssim 1\times 10^{-3}$ (down to about 100~mK). The XTRG 
truncation provides faithful estimate of error in the computed 
free energy, and the small $\epsilon$ value thus guarantee high accuracy of computed 
thermal data down to low temperature.

The XTRG simulations start from the initial density matrix 
$\rho_0(\tau)$ at a very high temperature $T\equiv 1/\tau$ 
(with the inverse temperature $\tau\ll 1$), represented in a 
matrix product operator (MPO) form~\cite{Chen2017}. The 
series of density matrices $\rho_n(2^n \tau)$ ($n\geq1$) at
lower temperatures are obtained by iteratively multiplying and 
compressing the MPOs $\rho_n = \rho_{n-1} \cdot \rho_{n-1}$.
As a powerful thermodynamic solver, XTRG has been 
successfully applied in solving triangular-lattice spin models
\cite{Chen2019TLH} and related compounds~\cite{Li2020TMGO,
Hu2020TMGO}, Kitaev model~\cite{Han2020} and materials
\cite{Li2021RuCl}, correlated fermions in ultra-cold quantum gas
\cite{Chen2021Hubbard}, and even moir\'e quantum materials
\cite{Lin2021exciton}.\\

\noindent{\textbf{Automatic parameter searching.}} 
By combining the thermodynamic solver XTRG and efficient 
Bayesian optimization approach, the optimal model parameters 
can be determined automatically via minimizing the fitting loss function 
between the experimental and simulated data, i.e.,
\begin{equation}
\mathcal{L}(\mathbf{x}) = \frac{1}{N_\alpha} \sum_{\alpha}
\frac{(O^\mathrm{exp}_\alpha-O^{\rm sim, \mathbf{x}}_\alpha)^2}
{(O^{\rm exp}_\alpha)^2}.   
\end{equation}
$O^{\rm exp}_\alpha$ and $O^{\rm sim, \mathbf{x}}_\alpha$ 
are respectively the experimental and simulated quantities with 
given model parameters $\mathbf{x} \equiv \{ J_{xy}, J_z, J_{\rm PD}, 
J_{\Gamma}, g_{ab,c}, \chi^{\rm vv}_{ab,c} \}$. The index $\alpha$ 
labels different physical quantities, e.g., magnetic specific heat 
and susceptibilities, and $N_\alpha$ counts the number of data 
points in $O_\alpha$. The optimization of $\mathcal{L}$ over the 
parameter space spanned by $\{J_{xy}, J_z, J_{\rm PD}, J_{\Gamma}\}$ 
is conducted via the Bayesian optimization~\cite{Yu2021}.
The Land\'e factor $g_{ab,c}$ and the Van Vleck paramagnetic 
susceptibilities $\chi^{\rm vv}_{ab,c}$ are optimized via the Nelder-Mead 
algorithm for each fixing $\{J_{xy}, J_z, J_{\rm PD}, J_{\Gamma}\}$. 
In practice, we perform the automatic parameter searching using 
the package QMagen developed by some of the authors~\cite{Yu2021,
QMagen}, and the results shown in the main text are obtained via over 
450 Bayesian iterations. After that, we introduce an additional parameter, 
the next-nearest-neighbor Heisenberg term $J_2$, and perform 
another 400 searching iterations. We find $J_2$ is indeed negligibly 
small and the obtained optimal parameters are robust.\\

\noindent{\textbf{Density matrix renormalization group.}}
The ground state magnetization curves of the easy-axis TLAF model 
for NBCP are computed by the density matrix renormalization group 
(DMRG) method~\cite{White1992}, which is a powerful variational 
algorithm based on the matrix product state ansatz. The DMRG 
simulations are performed on YC6$\times$15 lattice, and we retain 
bond dimension up to $D=1024$ with truncation error $\epsilon < 3 
\times 10^{-5}$, which guarantees well converged DMRG data.\\

\noindent{\textbf{Classical Monte Carlo simulations.}} 
We replace the $S=1/2$ operators by classical vectors, 
$S^{x,y,z}_i \to S\hat{n}_i$, where $\hat{n}_i$ is a unit vector, 
and $S=1/2$ is the spin quantum number. We use the standard 
Metropolis algorithm with single spin update. The largest system 
size is 48$\times$48. We compute the Binder ratio associated 
with the UUD-phase order parameter 
$\psi = m_1 + m_2\exp(i2\pi/3)+m_3\exp(-i2\pi/3)$, 
where $m_{1,2,3}$ are respectively the $S^z$-axis magnetization 
of the three sublattices, as well as the in-plane spin stiffness 
$\rho$~\cite{Stephan2000}. We locate the three-state Potts 
transition by examining the crossing of the Binder ratio, 
and the BKT transition by the criterion $\rho_c = (2/\pi) T_c$. 

{In the simulations, we use the natural unit in the calculation and thus 
the following process is required for comparing the model calculation results
in the natural unit to experimental data in SI units:
\begin{itemize}
\item[(1)] The value of temperature $T$ in natural unit should be 
multiplied by a factor of $J_{xy}$, and change it thus to the unit of Kelvin, 
where $J_{xy} = 0.88~{\rm K}$ is taken as the energy scale in the calculation.
\item[(2)] Multiply the value of specific heat $C_{\rm m}$ in natural unit by a factor of 
$R$, i.e., the ideal gas constant, and change it to the unit of J mol$^{-1}$ K$^{-1}$.
\item[(3)] Multiply the magnetic field $h$ in natural unit by a factor of 
${J_{xy} k_B}/{(g_z \mu_B)}$ and it is now in unit of {\rm Tesla},
where $g_z$ is the Land\'e factor along $S_z$ direction and $\mu_{B}$ the
Bohr magneton.
\end{itemize}
}

\noindent{\textbf{Sample preparation and susceptibility
measurements.}}
Single crystals of Na$_2$BaCo(PO$_4$)$_2$ were prepared by 
the flux method starting from Na$_2$CO$_3$ (99.9\%), 
BaCO$_3$ (99.95\%), CoO (99.9\%), (NH$_4$)$_2$HPO$_4$ (99.5\%), 
and NaCl (99\%), mixed in the ratio 2:1:1:4:5. Details of the heating 
procedure were given in Ref.~\cite{Zhong2019}. The flux generated 
after the reaction is removed by ultrasonic washing. The anisotropic 
magnetic susceptibility measurements in this work were performed 
using a SQUID magnetometer (Quantum Design MPMS 3). 
The magnetic susceptibility as a function of temperature was measured 
in zero field cooled runs. During the measurements, magnetic field of 
0.1 T was applied either parallel or perpendicular to the $ {c}$ axis. 
In the latter (in-plane) measurements, no  anisotropy is observed 
in the obtained susceptibility data. \\

\noindent{\bf{Data availability}}\\
The data that support the findings of this study are 
available from the corresponding author upon reasonable request.\\

\noindent{\bf{Code availability}}\\
All numerical codes in this paper are available 
upon request to the authors. \\

%  ====== Bib ======= %
% \bibliographystyle{apj}
%\bibliographystyle{naturemag}
%\bibliography{NBCPRef_Nat} 
% \bibliography{NBCPRef} 


$\,$\\
\textbf{Acknowledgements} \\

W.L. and Y.G. are indebted to Tao Shi for stimulating discussions, 
W.L. would also thank Xue-Feng Sun and Jie Ma for valuable 
discussions on the experiments. This work was supported by the 
National Natural Science Foundation of China 
(Grant Nos. 12222412, 11834014, 11874115, 11974036, 11974396, 12047503, and 12174068), 
the Strategic Priority Research Program of 
the Chinese Academy of Sciences (Grant No. XDB33020300),
and CAS Project for Young Scientists in Basic Research (Grant 
Nos. YSBR-057 and YSBR-059). 
We thank the HPC-ITP for the technical support and generous allocation of CPU time.

$\,$\\
\textbf{Competing interests} \\
The authors declare no competing interests. 

$\,$\\
\textbf{Author contributions} \\
W.L., Y.Q., and Y.W. initiated this work. Y.G. and H.L. performed 
XTRG and DMRG calculations of the TLAF model. Y.W.
conducted the symmetry and semi-classical analyses. X.T.Z., 
F.Y., and X.L.S. did the CEF point charge model analysis 
and DFT calculations. Y.C.F. undertook the MC  simulations. 
R.Z. prepared the sample and performed the susceptibility 
measurements. All authors contributed to the analysis of 
the results and the preparation of the draft. Y.W. and W.L. 
supervised the project. 

$\,$\\
\textbf{Additional information} \\
\textbf{Supplementary Information} is available in the online version of the paper. \\
\noindent
\textbf{Correspondence} and requests for materials should be addressed to Y.W. or W.L.

\clearpage
\renewcommand{\figurename}{\textbf{Supplementary Figure}}
\renewcommand{\tablename}{\textbf{Supplementary Tabel}}
\onecolumngrid
\setcounter{figure}{0}
\renewcommand\thefigure{\arabic{figure}}
\begin{center}
{\large Supplementary Information for}
$\,$\\
\textbf{\large{Spin supersolidity in nearly ideal easy-axis triangular quantum antiferromagnet Na$_2$BaCo(PO$_4$)$_2$}}

$\,$\\
Gao \textit{et al.}
\end{center}
% \date{\today}
\maketitle
\onecolumngrid
\setcounter{figure}{0}
\renewcommand\thefigure{\arabic{figure}}
\newpage

%=====================

\date{\today}

\setcounter{subsection}{0}
\setcounter{figure}{0}
\setcounter{equation}{0}
\setcounter{table}{0}

\renewcommand{\thesubsection}{\normalsize{Supplementary Note \arabic{subsection}}}
\renewcommand{\theequation}{\arabic{equation}}
\renewcommand{\thefigure}{\arabic{figure}}
\renewcommand{\thetable}{\arabic{table}}

\subsection{Crystal electric field calculations of 
Na$_2$BaCo(PO$_4$)$_2$} According to the Hund's rule, 
the lowest-energy electron structure of a free Co$^{2+}$ ion 
is $^4F(L =3, S = 3/2)$ with the high-spin state. We consider 
the energy spectrum of $^4F$ state under CEF and spin-orbit 
coupling with the effective Hamiltonian
\begin{equation}
H = H_ {\rm SOC} + H_{\rm CEF} = \lambda S \cdot L + 
\sum_{n, m} B_{n}^{m} O_{n}^{m},
\end{equation}
where $S$ is the spin angular momentum, $L$ is the orbital angular 
momentum, and $O_{n}^{m}$'s are the Stevens' operators with 
multiplicative CEF parameters  $B_{n}^{m}$. Due to the time-reversal 
symmetry constraint, the operator degree $n$ is required to be even,
and $m$ is the operator order which satisfies $-n \leq  m \leq n$. It is
necessary to treat both CEF and spin-orbit coupling non-perturbatively, 
namely the \textit{intermediate coupling scheme}, where $H_{\rm CEF}$ 
acts only on the orbital angular momentum $L$~\cite{Abragam1971}. 

\begin{table*}[tbp]
\centering
\caption{\label{cf_state} The two lowest CEF states in the Kramers doublet.}
\begin{ruledtabular}
\begin{tabular}{ccccccccccccccccccc}
\multicolumn{19}{c}{J = 9/2} \\
\colrule
$L_z$    & \multicolumn{2}{c}{-3} & \multicolumn{3}{c}{-2} & \multicolumn{3}{c}{-1} & \multicolumn{2}{c}{0} & \multicolumn{3}{c}{1} & \multicolumn{3}{c}{2} & \multicolumn{2}{c}{3} \\
\colrule
$S_z$  & -1/2& 1/2 & -3/2& -1/2& 3/2 & -3/2& 1/2 & 3/2 & -1/2& 1/2 & -3/2& -1/2& 3/2 & -3/2& 1/2 & 3/2 & -1/2& 1/2 \\
\colrule
1     & -0.41 & -0.02 & -0.71 & -0.03 & -0.04 & -0.01 & -0.19 & -0.02 & -0.32 & -0.04 & -0.21 & -0.02 & 0.10 & -0.01 & 0.29 & 0.08  & 0.19 & 0.05 \\
2     & 0.05 & -0.19 & 0.08  & -0.29 & 0.01 & -0.10 & 0.02 & -0.21 & 0.04 & -0.32 & 0.02 & -0.19 & -0.01 & -0.05 & -0.03 & 0.71 & -0.02 & 0.41 \\
\end{tabular}%
\end{ruledtabular}
\end{table*}%

Constructing a point charge model directly from 
Na$_2$BaCo(PO$_4$)$_2$ structure and perform the calculations 
with the open-source package PyCrystalField~\cite{pycef}, we obtain 
the three CEF constants $B_2^0 = -3.75$~meV, $B_4^0 = 0.45$~meV, 
and $B_4^3 = 11.51$~meV. Considering the spin-orbit coupling constant 
$\lambda = -65.34$~meV from experiments~\cite{soc-const,atomdatabase}, 
the obtained CEF levels are shown in Supplementary Figure~\ref{FigM1}, where the two 
lowest levels constitute a Kramers doublet, i.e., the effective spin-1/2 
degree of freedom in the spin-orbit magnet. We further show the 
wavefuctions of the two CEF states with coefficients of each 
$(L_z, S_z)$ components listed in Supplementary Table~\ref{cf_state}, 
where the components with large $J_z$, like $\pm9/2$, $\pm7/2$, 
etc., have relatively large weights. This suggests an easy-axis 
anisotropy of the compound from the single-ion physics. 
Meanwhile, the CEF splitting between the lowest Kramers doublet 
to the higher levels is about 900 K ($\sim$ 71 meV), rendering clearly 
an effective spin-1/2 magnet at relevant temperature in the study 
of spin supersolidity in this work.\\

% //CEF level of 3d7 Co ion
\begin{figure}
\includegraphics[width=8cm]{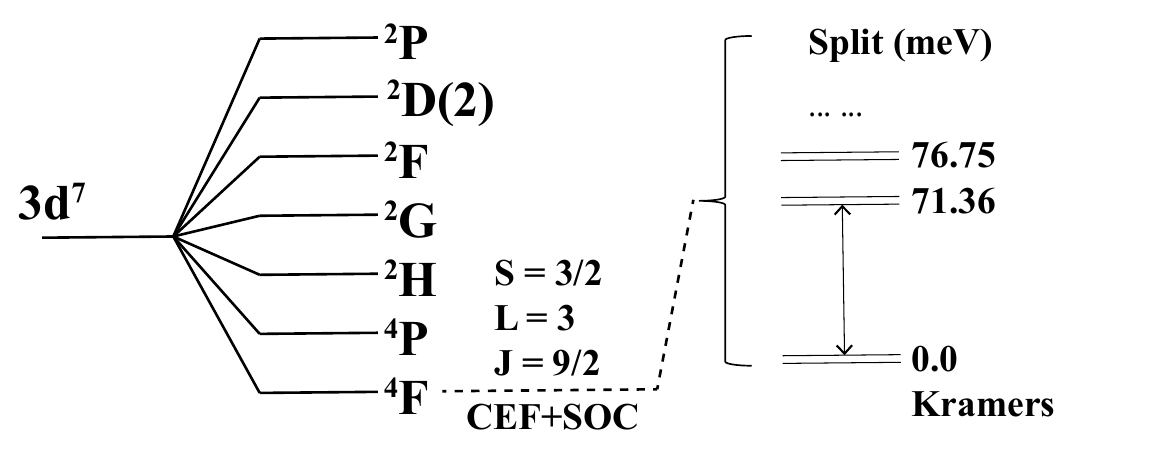}
\caption{The electronic state of Co$^{2+}$ ion in the compound
Na$_2$BaCo(PO$_4$)$_2$, where the two degenerate lowest 
levels form an effective spin-1/2 doublet.}
\label{FigM1}
\end{figure}

\subsection{DFT+U calculations of Na$_2$BaCo(PO$_4$)$_2$}
Here we employ the density functional theory (DFT)+U approach
to estimate the spin exchange between Co$^{2+}$ ions. We use 
the experimental lattice constants~\cite{Zhong2019} in our DFT+U  
calculations~\cite{Anisimov1991,LDAU,dudarev1998} with the 
Perdew-Burke-Ernzerhof~\cite{PBE1996} functional to evaluate 
the spin couplings. 

First we construct a series of magnetic configurations on the 
three sublattices formed by Co$^{2+}$ ions in Supplementary Figure~\ref{FigS:DFT}. 
We compute their total energies in different sizes of supercells and 
list the results in Supplementary Table~\ref{tab:DFT-energy}. From the results, 
we find the ferromagnetic and interlayer antiferromagnetic (AFM) 
states have very close energies, suggesting a very weak interlayer 
coupling (estimated as $J_\perp\simeq0.05$~K). In the calculations 
with $2\times2\times1$ supercell, the total energy of the ferrimagnetic 
state is about 0.3~eV lower than that of ferromagnetic state, and the 
corresponding nearest neighbor (NN) coupling $J_1$ is estimated to 
be about 30 K, which is clearly larger than the results in the main text 
(and also certain previous estimation~\cite{Li2020}). In addition, the three 
stripe states have almost the same energy, as determined from computing 
the energy difference between the stripe and ferromagnetic states.
These so-obtained exchange coupling is much stronger than the model
in the main text, such inconsistency reflects the essential challenges 
of determining spin couplings between 3$d$ ions from DFT+U calculations.

\begin{figure}
\includegraphics[width=12cm]{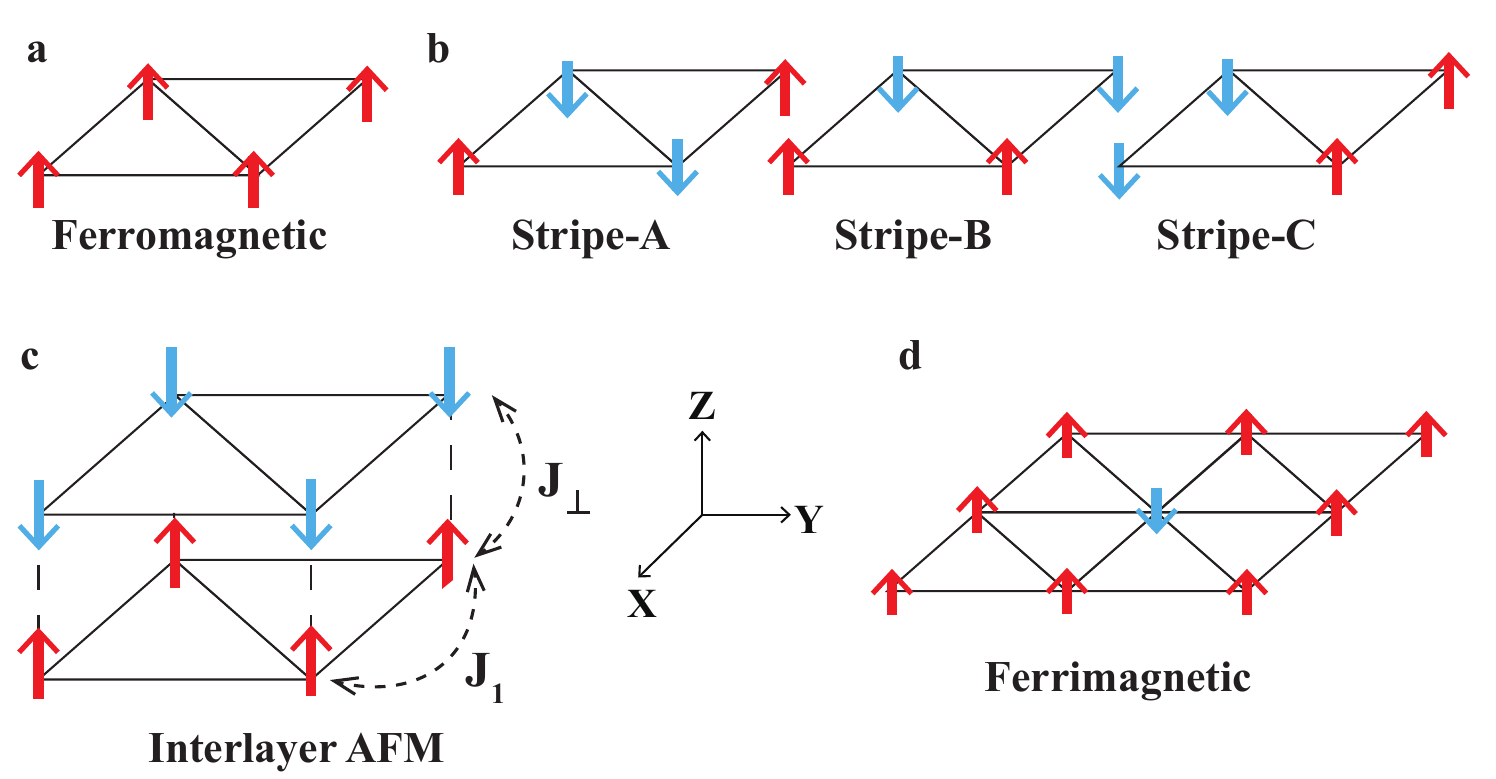}
\caption{\label{FigS:DFT} \textbf{a}-\textbf{d} show the sketches of 
different magnetic configurations. The energies of \textbf{a} ferromagnetic 
and \textbf{b} stripe states are also calculated in enlarged supercells 
in order to compare with other states in \textbf{c} and \textbf{d} directly
(see Supplementary Table~\ref{tab:DFT-energy}).}
\end{figure}

\begin{table}[tbp]
    \caption{\label{tab:DFT-energy}
    Magnetic space group (MSG) numbers and energies of different configurations, calculated by DFT+U method.
    }
    \begin{ruledtabular}
    \begin{tabular}{lcccr}
    \textrm{Magnetic Configuration}&
    \textrm{MSG Number}&
    \textrm{Supercell Size}&
    \textrm{Energy/eV}&
    \textrm{Difference/eV}\\
    \colrule
    Ferromagnetic &164.89	& $1\times1\times2$ &-184.90846 & ----- \\
    Interlayer-AFM&165.96	& $1\times1\times2$ &-184.90828 & 0.00018\\
    \colrule
    Ferromagnetic&164.89	& $2\times2\times1$	&-368.29838& 0.00000\\
    Ferrimagnetic&164.89	& $2\times2\times1$	&-368.59576& -0.29738\\
    AFM stripe-A&14.83	& $2\times2\times1$	&-368.30225& -0.00387\\
    AFM stripe-B&14.83	& $2\times2\times1$	&-368.30227& -0.00389\\
    AFM stripe-C&14.83	& $2\times2\times1$	&-368.30226& -0.00388\\
    \end{tabular}
    \end{ruledtabular}
\end{table}

In Supplementary Figure~\ref{FigS:Band} we adopt an alternative way to estimate the spin 
exchange couplings~\cite{t2u,Wellm2021}. In Supplementary Figure~\ref{FigS:Band}\textbf{a} 
the orbital projected band structure of Na$_2$BaCo(PO$_4$)$_2$ with 
Hubbard $U_{\rm eff}= 2$~eV is shown. We see near the Fermi energy 
are mainly 3$d$ electron bands, well separated from the other bands.
We choose the 3$d$ orbitals as the bases of Wannier functions to 
compute the major hopping amplitudes between the near neighboring 
Wannier centers, and the exchange coupling $J_1$ between a pair of 
Co$^{2+}$ ions can be estimated as $J_1=(16/81)t^2/U$~\cite{Wellm2021}. 
The resulting $J_1$ with $U_{\rm eff}$ values below 2.5~eV are plotted in 
Supplementary Figure~\ref{FigS:Band}\textbf{b}, from which we find that $J_1$ decreases 
rapidly as $U_{\rm eff}$ increases. For $U_{\rm eff}$ above 2.5~eV, 
the Co 3$d$-orbitals are mixed with the oxygen 2$p$-orbitals, and the 
simple formula for $J_1$ estimation becomes no longer applicable. 
Therefore we perform a (second-order) polynomial fit of the DFT results 
up to $2.5$~eV, and extrapolate to large $U_{\rm eff}$. We find $J_1$ 
becomes about 2.4~K for $U_{\rm eff} = 5$~eV, in agreement with the 
energy scales of determined spin exchange in the main text. 

% ======== DFT + U ======== %
\begin{figure}
\includegraphics[width=14cm]{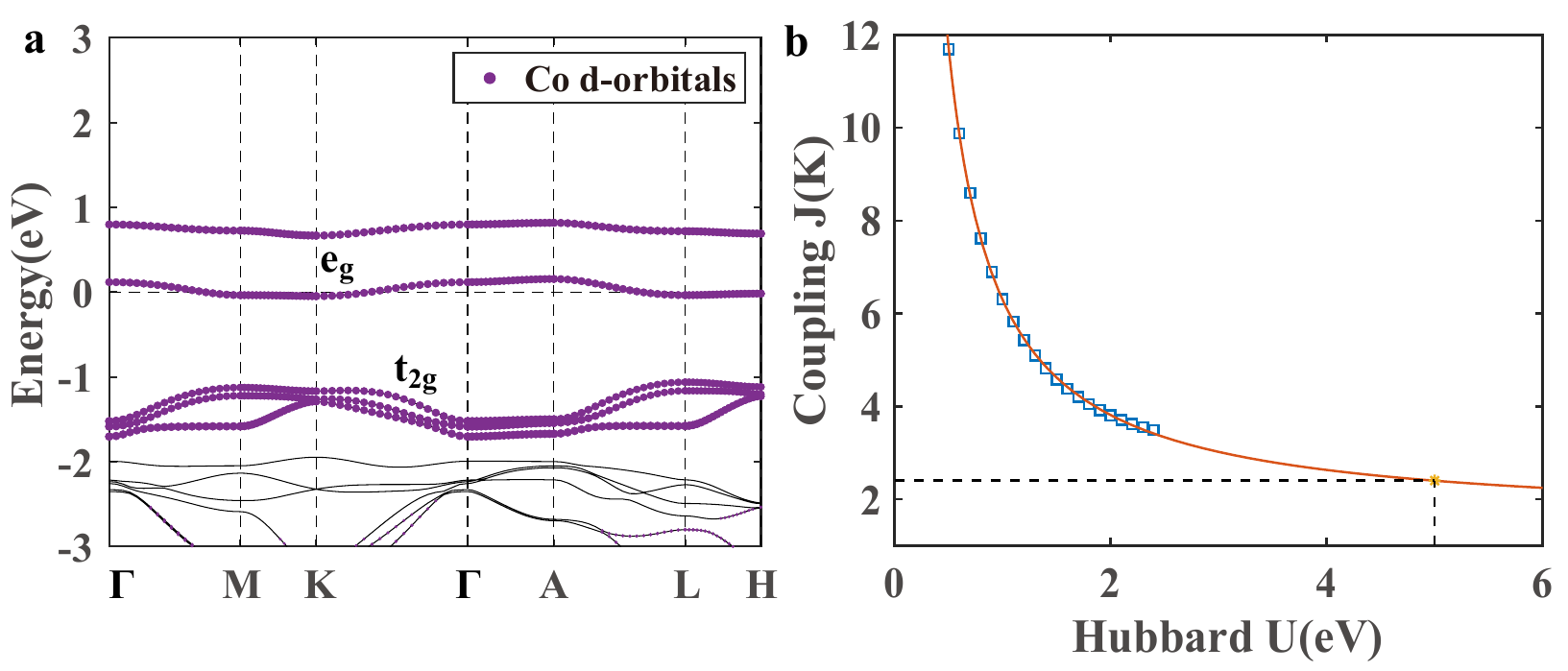}
\caption{\label{FigS:Band}\textbf{a} The Co 3$d$-orbital projected 
band structure of Na$_2$BaCo(PO$_4$)$_2$. \textbf{b} The estimated 
exchange strengths $J_1$ drawn from the Wannier functions are plotted 
vs. the Hubbard $U$. The blue dots are the Wannier functions results, 
and the red line represent the second-order polynomial extrapolation.}
\end{figure}

Nevertheless, note the true exchange path is a super-super exchange 
through the Co-O-O-Co path shown below instead of the direct Co-Co 
exchange. In Supplementary Figure~\ref{FigS_eleden} we provide the charge density 
contour obtained by DFT+U calculations, which visualize the spin 
exchange path. The minimum charge density in the Co-O-O-Co path in 
Supplementary Figure~\ref{FigS_eleden}\textbf{a} is about $0.03~{\rm e}~ {\rm Bohr}^{-3}$. 
The out-of-plane contour map in Supplementary Figure~\ref{FigS_eleden}\textbf{b} 
shows the very weak overlap of charge density distributions between 
two adjacent triangular planes, and the minimum charge density is 
about $0.004~{\rm e}~ {\rm Bohr}^{-3}$ in the supposed super-super-super 
exchange path indicated by the dash line. 

\begin{figure}
\includegraphics[width=15cm]{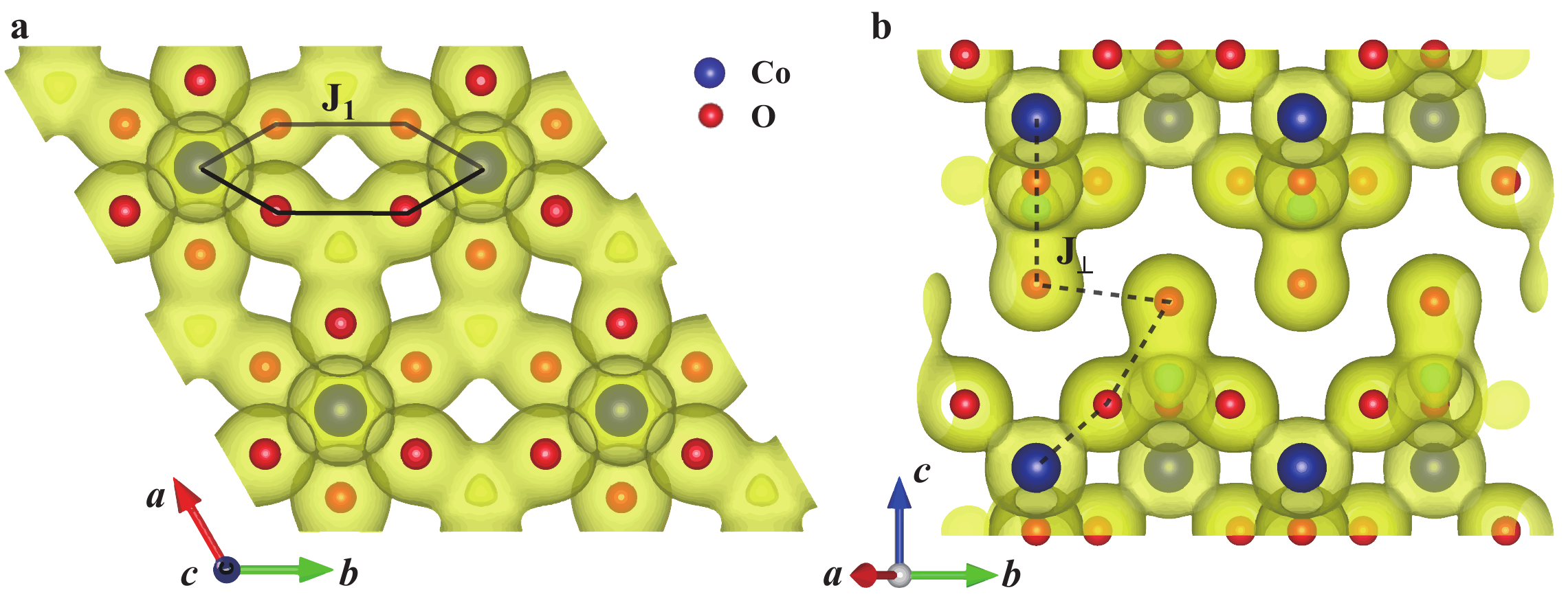}
\caption{The intra- and inter-plane charge density contour maps. 
The red and blue dots represent respectively the O and Co ions. 
The solid lines in \textbf{a} denote the super-super exchange path 
between a pair of near neighbor Co ions within the same triangular 
plane. The dashed lines in \textbf{b} denote the possible super-super-super
exchange path of two Co ions in adjacent planes.}
\label{FigS_eleden}
\end{figure}

\subsection{Crystal symmetry analysis and the effective 
spin Hamiltonian}
In this section, we use the symmetries of 
the material to constrain the possible exchange interactions 
between two neighbor Co$^{2+}$ ions.
\begin{itemize}
\item The Co$^{2+}$ ions occupy the Wychoff position $1b$ of 
the space group $P\bar{3}m1$. Its site symmetry group is 
$\bar{3}m$ (see, Supplementary Figure~1\textbf{a} of the main text), 
which is generated by a 3-fold rotation w.r.t. the crystallographic 
$ {c}$ axis, a two-fold rotation w.r.t. the nearest neighbor (NN)
bond, and the inversion.
\item The center of two neighboring Co$^{2+}$ ions in the same 
basal planes corresponds to the Wyckoff position $3f$ with site 
symmetry group $2/m$. The site symmetry group is generated by 
a two-fold rotation w.r.t. the NN bond and the inversion.
\end{itemize}

We consider the exchange interactions between magnetic 
ions in the same basal plane. There are three translation-inequivalent 
NN bonds, which are related by the three-fold rotations.
Therefore it is sufficient to determine the exchange interaction on 
one bond and obtain the interactions on the other two by rotations.

Let $i$, $j$ denote two NN sites of Co$^{2+}$. Let $\hat{n}_{ij}$ be the unit vector that points form site $i$ to site $j$ 
and $\hat{\epsilon}_{ij} \equiv {c} \times \hat{n}_{ij}$. The most general bilinear exchange interaction between these two 
sites reads: $H_{ij} = \sum_{\alpha\beta}
J^{\alpha\beta}S_i^{\alpha} S_j^{\beta} $, where $\alpha, ~\beta $ run over  $\{ \hat{n}_{ij}, 
\hat{\epsilon}_{ij}, {c}\}$. Now consider all the symmetries that
preserve the bond $ij$. The inversion symmetry w.r.t. the center 
of the bond implies that $J^{\alpha\beta} = J^{\beta\alpha}$.
Furthermore, the 2-fold rotation w.r.t. the bond itself implies 
$J_{n\epsilon}=J_{nc}=0$. We are left with four independent, 
symmetry-allowed interactions 
$H_{ij} = J_H \textbf{S}_i\cdot\textbf{S}_j 
 + J_I({c}\cdot\textbf{S}_i) ({c}\cdot\textbf{S}_j)
 + J_{nn}(\hat{n}_{ij}\cdot \textbf{S}_i)(\hat{n}_{ij}\cdot \textbf{S}_j) 
 + J_{\epsilon c} [({c}\cdot\textbf{S}_i)(\hat{\epsilon}_{ij}\cdot \textbf{S}_j) 
 + ({c}\cdot\textbf{S}_j)(\hat{\epsilon}_{ij}\cdot \textbf{S}_i)]$. We recognize the first term as the 
Heisenberg exchange, the second Ising, the third pseudo-dipolar, 
and the last ``symmetric off-diagonal'' exchange interaction.

To align with the convention in previous works
\cite{Li2015YMGO2,LiYD2016,Zhu2018}, we recast 
the Hamiltonian via the following transformation: $J_{xy} = 
J_H + \frac{1}{2} J_{nn}$, $J_{z} =J_H+J_I$, $J_{\rm PD} = 
\frac{1}{4} J_{nn}$, $J_{\Gamma} = J_{n\epsilon}$.
We note, through such transformation, the parameters become $
J_{\pm \pm} \equiv  J_{\rm PD} $ and $J_{z\pm}\equiv J_{\Gamma} $ 
as adopted in Refs.~\cite{Li2015YMGO2,LiYD2016,Zhu2018}.
{Here we define $S^x$ and $S^z$, i.e., respectively spin 
(1,0,0) and (0,0,1) directions, as along the $a-$ and $c-$ axes},
and arrive at the Hamiltonian in Eq.~(1) of the main text.

{Another way to set up the coordinate system is taking the spin $(1,1,1)$ 
direction parallel to $c$-axis, and spin $(-1,-1,2)$ direction parallel to $a$-axis. 
Thus the 3-fold rotation operation is a cyclic permutation of three components
of the spin operator. For $\hat{n}_{ij} = \frac{1}{\sqrt{6}}(-1,-1,2)$, i.e., along the
$a$-axis, one has
\begin{equation}
\begin{split}
H_{ij} =& J_H \textbf{S}_i\cdot\textbf{S}_j  + \frac{J_I}{3}(S_i^x + S_i^y + S_i^z)(S_j^x + S_j^y + S_j^z)
+\frac{J_{nn}}{6}(-S_i^x - S_i^y + 2 S_i^z)(-S_j^x - S_j^y + 2 S_j^z) \\
&+\frac{J_{\epsilon c}}{6}\{(S_i^x + S_i^y + S_i^z)(S_j^x - S_j^y) + (S_i^x - S_i^y)(S_j^x + S_j^y + S_j^z)\}.
\end{split}
\end{equation}
The $J_{nn}$ and $J_{\epsilon c}$ terms can lead to a bond-dependent, Kitaev-type, 
interaction in the system, which, however, are found to be negligible in our spin model
established in the main text.
}
% =========== More XTRG Results =========== %
\subsection{XTRG results of TLAF model under nonzero fields}
Besides the zero-field magnetic specific heat $C_{\rm m}$ presented 
in the main text, here we present comparisons between the simulated 
$C_{\rm m}$ under various magnetic fields along the $c$ axis with the 
experimental results. As shown in Supplementary Figure~\ref{Fig:CmHF}, the model 
calculations and experiments are in very good agreement whenever 
the two experimental curves~\cite{Zhong2019, Li2020} coincide. 
In the lower temperature range, our simulated data can well 
reproduce the peak positions in quantitative agreement with 
experiment (c.f., Supplementary Figure~\ref{Fig:CmHF}\textbf{a, b}). 

In practical calculations, we perform XTRG calculations on two 
different lattice geometries (c.f. Fig~\ref{Fig:Geo}), i.e., YC$4\times6$ 
(used mainly in the automatic parameter searching) and YC$6\times9$ 
(larger-size calculations for validation). As shown in 
Fig~\ref{Fig:CmHF}\textbf{a}, above $T_{\text{cut}}= 1$~K, 
no significant difference between the two simulated data with 
Lattice~1 and 2 can be observed.

% ====== Magnetic Specific Heat under nonzero Field ====== %
\begin{figure}[h]
\includegraphics[width=0.9 \linewidth]{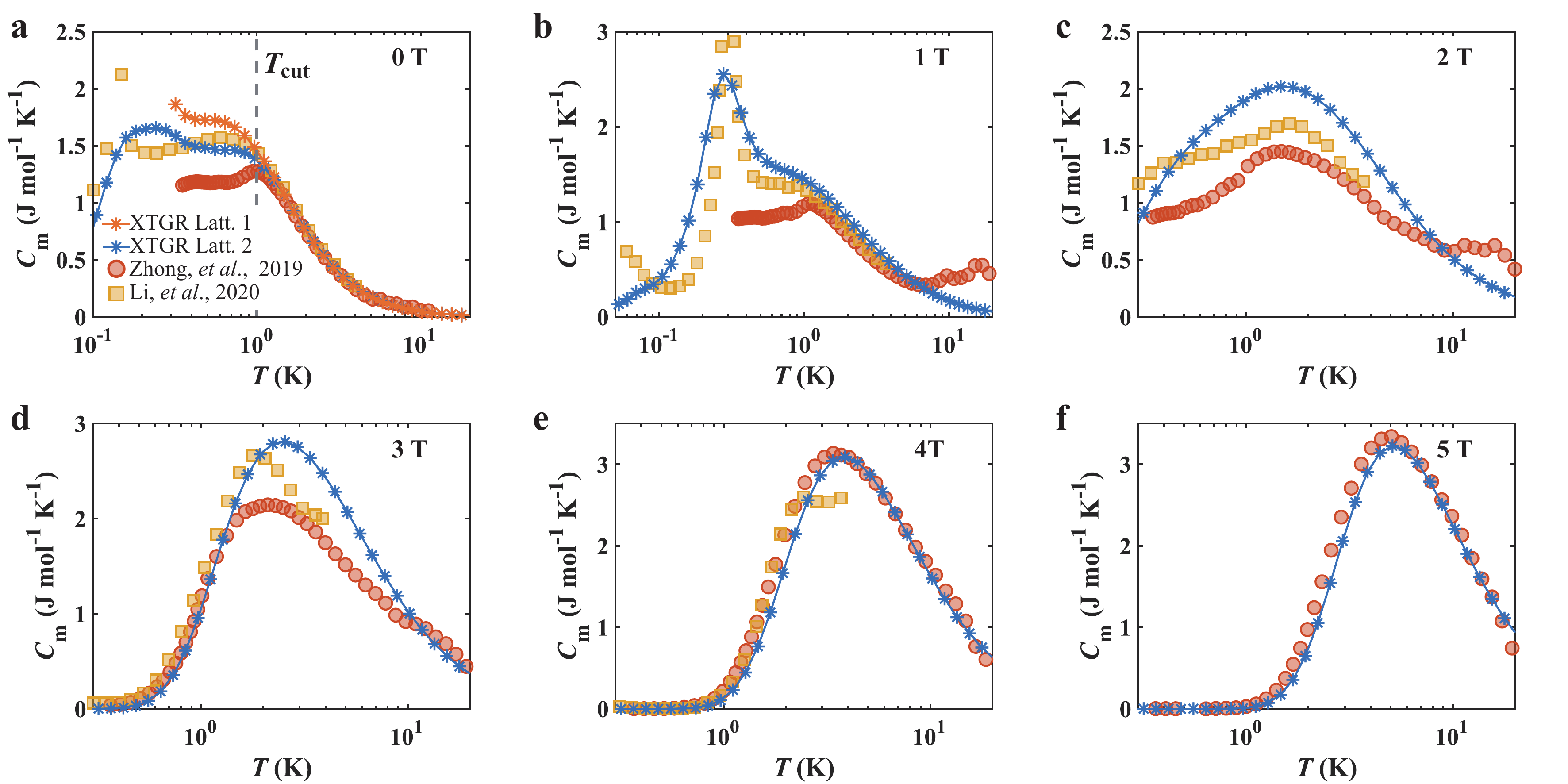}
\caption{\textbf{a}-\textbf{f} The magnetic specific heat of XTRG 
simulations under various fields along the $c$ axis, and comparisons 
to the experimental data (Zhong, \textit{et al.}, 2019~\cite{Zhong2019}, 
and Li, \textit{et al., 2020}~\cite{Li2020}). The simulations are 
performed on YC$4\times6$ (Lattice 1) and YC$6\times9$ 
(Lattice 2, see Supplementary Figure~\ref{Fig:Geo}).
} 
\label{Fig:CmHF}
\end{figure}

% // Lattice geometry //
\begin{figure}[h]
\includegraphics[width=8cm]{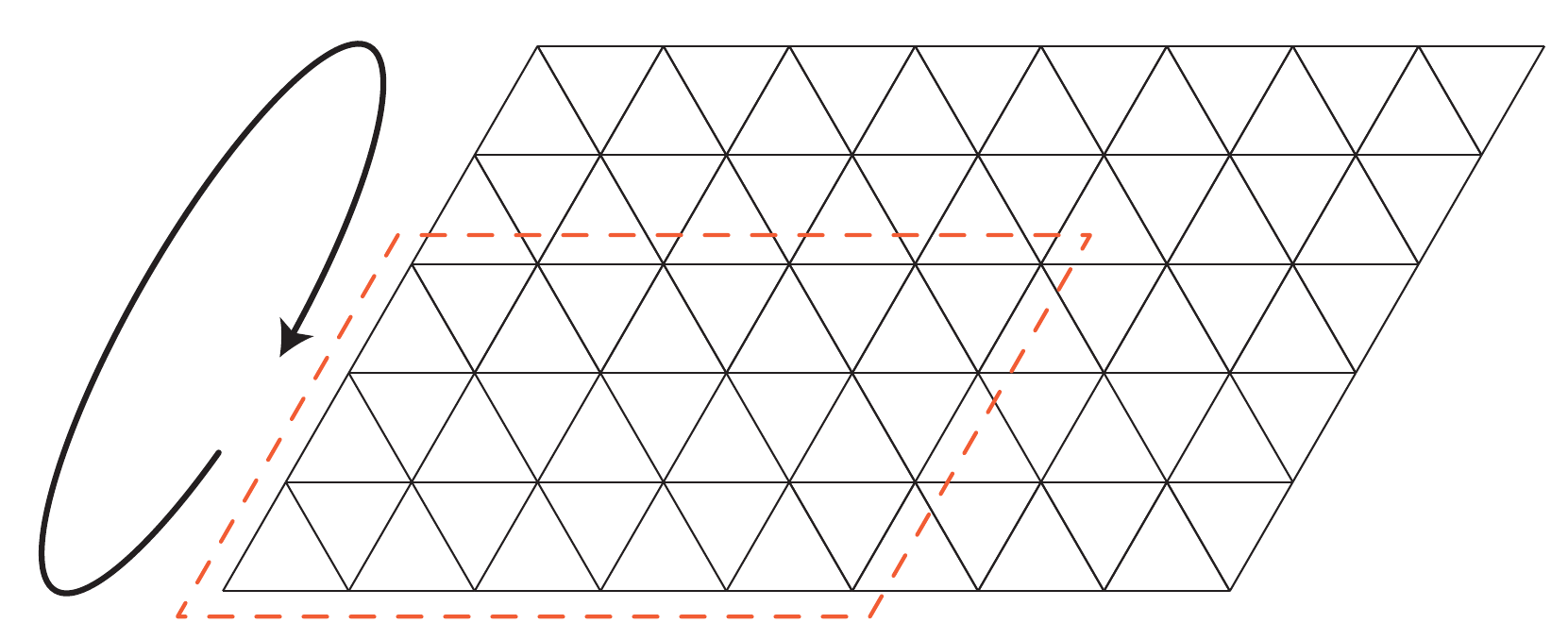}
\caption{The YC$6\times9$ and YC$4\times6$ (within the orange 
dash line) lattice geometries that are used in XTRG simulations.
The arrow indicates the periodic boundary condition along the
circumference of the cylinder.
} 
\label{Fig:Geo}
\end{figure}
% ============================================= %

% ======= Semiclassical analysis of transition fields ====== %
\subsection{Semi-Classical analysis of the TLH model under 
out-of-plane fields} 
It is known that the classical ground state of the TLAF model 
shows a three-sub-lattice structure. In the presence of a magnetic field, 
the model shows a sequence of phase transitions. Nevertheless, 
the said three-sub-lattice structure is preserved.

Suppose the interactions $J_{xy}$ and $J_{z}$ are dominant. It is then
natural to assume the subdominant interactions $J_{\rm PD}$ and
$J_{\Gamma}$ do not change the three-sub-lattice structure — in other
words, the magnetic unit cell remains to be $\sqrt{3}\times\sqrt{3}$. We 
now show that the classical magnetic phase diagram will be independent of $J_{\rm PD}$ 
and $J_{\Gamma}$.

Let $\bf{S}_{A,B,C}$ donate the classical spin vector in the sub-lattices 
A, B, and C, respectively. The classical energy reads:
\begin{equation}
\begin{split}
H =& \frac{N}{3} (J_a^{\alpha\beta} + J_b^{\alpha\beta} + J_c^{\alpha\beta}) S_A^{\alpha} S_B^{\beta}
	+(B, C) + (C,A) 
	- \frac{N \mu_0}{3} g_{\alpha\beta}B^{\alpha}(S_A^{\beta} + S_B^{\beta} + S_C^{\beta})\\
   =& N(J_{xy}S_A^{x} S_B^{x} + J_{xy}S_A^{y} S_B^{y} + J_{z}S_A^{z} S_B^{z}) 
	+(B, C) + (C,A) 
	- \frac{N \mu_0}{3} g_{\alpha\beta}B^{\alpha}(S_A^{\beta} + S_B^{\beta} + S_C^{\beta}),
\end{split}
\label{EqS:CEnergy}
\end{equation}
where $N$ is the number of lattice sites, $g_{\alpha \beta}$ is the general Land\'e factor, and $B^{\alpha}$ is the magnetic field along $\alpha$ direction. We see that the contributions 
form $J_{\rm PD}$ and $J_{\Gamma}$ cancel. An immediate consequence is that the classical ground states show an accidental U(1) symmetry w.r.t. the $z$ axis, which will be lifted by quantum fluctuations through the order by disorder mechanism.

We now review the classical magnetic ground states of the TLAF model
with easy-axis anisotropy, i.e., $J_z > J_{xy} > 0$. As the out-of-plane
field $B \parallel c$ increases, the model shows a sequence of four magnetic 
phases: the Y state, the up-up-down state, the V state, and the fully 
polarized state. These phase are separated by three critical fields, 
which we label $B_{c1}$, $B_{c2}$, and $B_{c3}$,

We would like to determine the values of these critical fields. 
Beginning with $B_{c3}$, let us consider the stability of the 
polarized phase. We write:
\begin{equation}
S_A^x = \sqrt{S} x_A,~~~S_A^y=\sqrt{S}y_A,~~~S_A^z=
S-\frac{x_A^2 + y_A^2}{2}.
\label{EqS:PL}
\end{equation}
The other two spins are written in the same manner. Substituting
Eq.~(\ref{EqS:PL}) above into the expression of energy
[Eq.~(\ref{EqS:CEnergy})] and expand to the quadratic order:
\begin{equation}
H = \frac{N}{2}(x^{\text{T}}M_x x + y^{\text{T}}M_y y),
\end{equation}
where $x = (x_A, x_B, x_C) ^ {\text{T}}$ and $y = (y_A, y_B, y_C)^{\text{T}}$ . The Hessian matrix:
\begin{equation}
M_x = M_y = 
\left(
\begin{matrix}
\frac{g_z\mu_0B}{3} - 2 J_z S & J_{xy}S & J_{xy}S\\
J_{xy}S & \frac{g_z\mu_0B}{3} - 2 J_z S &J_{xy}S\\
J_{xy}S&J_{xy}S&\frac{g_z\mu_0B}{3} - 2 J_z S 
\end{matrix}
\right)
\end{equation}
The three eigenvalues are:
\begin{equation}
\lambda_{1,2} = \frac{g_z\mu_0B}{3} -2 J_z S - J_{xy} S,~~~
\lambda_3 =\frac{g_z\mu_0B}{3} -2 J_z S + 2 J_{xy} S.
\end{equation}
The stability of polarized state requires $\lambda_{1,2} \geq 0$, 
which implies:
\begin{equation}
	g_z\mu_0 B_{c3} = 3(J_{xy} + 2J_z) S.
\end{equation}

We then determine $B_{c1}$ and $B_{c2}$. To this end, consider the stability of the UUD phase.
We write:
\begin{equation}
\begin{split}
S_A^x = \sqrt{S} x_A,~~~S_A^y &=-\sqrt{S}y_A,~~~
S_A^z =-S+\frac{x_A^2 + y_A^2}{2};\\
S_{B,C}^x = \sqrt{S} x_{B,C},~~~S_{B,C}^y &=\sqrt{S}y_{B,C},~~~
S_{B,C}^z=S-\frac{x_{B,C}^2 + y_{B,C}^2}{2}.
\end{split}
\end{equation}
The energy is given by:
\begin{equation}
H = \frac{N}{2}(x^{\text{T}}M_x x + y^{\text{T}}M_y y),
\end{equation}
where the Hessian matrices:
\begin{equation}
M_x  = 
\left(
\begin{matrix}
2 J_z S - \frac{g_z\mu_0B}{3}  & J_{xy}S & J_{xy}S\\
J_{xy}S & \frac{g_z\mu_0B}{3} &J_{xy}S\\
J_{xy}S&J_{xy}S&\frac{g_z\mu_0B}{3}  
\end{matrix}
\right),
M_y = 
\left(
\begin{matrix}
2 J_z S - \frac{g_z\mu_0B}{3}  &- J_{xy}S & -J_{xy}S\\
-J_{xy}S & \frac{g_z\mu_0B}{3} &J_{xy}S\\
-J_{xy}S&J_{xy}S&\frac{g_z\mu_0B}{3}  
\end{matrix}
\right).
\end{equation}
The eigenvalues of $M_x$ and $M_y$ are identical. They are given by:
\begin{equation}
\lambda_1 = \frac{g_z\mu_0B}{3} - J_{xy},~~~
\lambda_{2,3} = J_z S + \frac{J_{xy} S}{2}
\pm \sqrt{(J_zS-\frac{g_z\mu_0B}{3}-\frac{J_{xy} S}{2})^2+2(J_{xy}S)^2}.
\end{equation}
The stability condition requires:
\begin{equation}
\frac{g_z\mu_0B}{3} - J_{xy} \geq 0,
~~~(J_z S + \frac{J_{xy}S}{2})^2 \geq
(J_zS-\frac{g_z\mu_0B}{3}-\frac{J_{xy} S}{2})^2+2(J_{xy}S)^2
\end{equation}
We deduce:
\begin{eqnarray}
& g_z\mu_0B_{c1} = 3 J_{xy} S, \\
& g_z\mu_0B_{c2} = 3(J_z - \frac{J_{xy}}{2}+\sqrt{J_z^2+J_zJ_{xy}-\frac{7}{4}J_{xy}^2})S.
\end{eqnarray}

\end{document}